\documentclass[aps,prd,longbibliography,nofootinbib]{revtex4-2}

\usepackage{graphicx}
\usepackage{amsmath,amssymb,bm}
\usepackage{booktabs}
\usepackage{microtype}
\usepackage{natbib}
\usepackage{hyperref}
\usepackage{multirow}
\usepackage{placeins}
\usepackage{xcolor}
\usepackage{xurl}
\hypersetup{colorlinks=true,allcolors=blue,hypertexnames=false}
\usepackage{eurosym}

\graphicspath{{./}{figs/}}
\renewcommand{\arraystretch}{1.12}

\newcommand{\SI}[2]{\ensuremath{#1\,\text{#2}}}
\newcommand{\SIrange}[3]{\ensuremath{#1\text{--}#2\,\text{#3}}}
\newcommand{\percent}{\%}

\newcommand{\OGS}{\ensuremath{\mathrm{OGS}}}
\newcommand{\LEO}{\ensuremath{\mathrm{LEO}}}

\newcommand{\GEO}{\ensuremath{\mathrm{GEO}}}
\newcommand{\SDA}{\ensuremath{\mathrm{SDA}}}
\newcommand{\CCSDS}{\ensuremath{\mathrm{CCSDS}}}
\newcommand{\RF}{\ensuremath{\mathrm{RF}}}
\newcommand{\AO}{\ensuremath{\mathrm{AO}}}
\newcommand{\ATP}{\ensuremath{\mathrm{ATP}}}
\newcommand{\DTE}{\ensuremath{\mathrm{DTE}}}
\newcommand{\Gbps}{\ensuremath{\mathrm{Gb\,s^{-1}}}}
\newcommand{\Mbps}{\ensuremath{\mathrm{Mb\,s^{-1}}}}
\newcommand{\TB}{\ensuremath{\mathrm{TB}}}

\newcommand{\um}{\ensuremath{\mathrm{\mu m}}}

\newcommand{\costTB}{c_{\mathrm{TB}}}
\newcommand{\Qyr}{Q_{\mathrm{yr}}}
\newcommand{\Cnet}{C_{\mathrm{net}}}
\newcommand{\Rline}{R_{\mathrm{line}}}
\newcommand{\Hday}{H_{\mathrm{day}}}
\newcommand{\Avail}{A_{\mathrm{net}}}
\newcommand{\score}{S}

\begin{document}

\title{Optical Ground Stations for Space Communications:\\
Systems Engineering, Availability, and Service Economics Through 2030}

\author{Slava G. Turyshev}
\affiliation{
Jet Propulsion Laboratory, California Institute of Technology,\\
4800 Oak Grove Drive, Pasadena, CA 91109-0899, USA
}%

\date{\today}

\begin{abstract}
Optical ground stations (OGSs) are becoming networked infrastructure for high-rate space-to-Earth communications, but their adoption is governed by service availability and utilization as much as by optical line rate. This paper develops a systems-engineering and service-economics assessment of the OGS sector as of June~2026. The analysis combines public flight demonstrations and operational records with scalar link-budget, availability, and cost-normalization models. Public benchmarks span \SI{25}{\Mbps} from interplanetary range, \SI{260}{\Mbps}-class lunar links, \SI{1.2}{\Gbps}-class ISS relay, \SI{1.8}{\Gbps} operational GEO relay, \SI{120}{\Gbps}-class direct-to-ground demonstrations in China, and \SI{200}{\Gbps} from LEO in NASA's TBIRD mission. The resulting conclusion is that the bottleneck has shifted from peak line rate to repeatable service under weather, acquisition, scheduling, and operations constraints. Under one explicit planning normalization---a \SI{10}{\Gbps} near-Earth station, annualized cost of \$2~million/year, scheduled pre-weather optical contact time of 0.5~h/day, and weather-inclusive combined efficiency $\eta=0.7$---the fixed-cost component is of order $3\times10^{3}$--$4\times10^{3}$~USD/TB. This number is a sensitivity anchor, not a tariff forecast; the controlling variables are duty factor, effective weather diversity, shared-network loading, and service-level allocation. The public industrial evidence is best interpreted as a stratified value chain, not as a single vendor ranking. The defensible 2030 baseline is hybrid optical+radio-frequency (RF): optical for throughput, relay, and spectrum relief; RF for continuity, contingency, and assured command paths.
\end{abstract}


\maketitle


\section{Introduction}
\label{sec:intro}

An optical ground station (OGS) is the terrestrial endpoint of a free-space optical link between a spacecraft and Earth. Functionally, it is the optical analogue of a radio-frequency (RF) ground station: it acquires and tracks the spacecraft, transmits and/or receives highly collimated optical beams, mitigates atmospheric effects, performs modulation/demodulation and forward-error-correction (FEC) functions, and hands traffic into terrestrial networks. The engineering motivation follows from wavelength. For a given transmit aperture, the diffraction-limited beam width scales as $\lambda/D$, so optical carriers near \SI{1{,}550}{nm} provide much narrower beams and higher data concentration than microwave carriers for comparable spacecraft size, weight, and power \cite{NASA_SST_OGS_2025,NASA_OGS1_2024}. Table~\ref{tab:abbrev} lists the abbreviations used throughout the paper.

\begin{table}[htbp]
\centering
\caption{Abbreviations used in the manuscript. Organizational names are expanded for clarity; several mission and product names are retained as proper names after expansion.}
\label{tab:abbrev}
\setlength{\tabcolsep}{4pt}
\renewcommand{\arraystretch}{1.12}
\begin{tabular}{@{}p{0.10\textwidth}p{0.34\textwidth}p{0.10\textwidth}p{0.34\textwidth}@{}}
\toprule
Abbrev. & Meaning & Abbrev. & Meaning \\
\midrule
AO & Adaptive optics & APD & Avalanche photodiode \\
ATP/PAT & Acquisition/tracking/pointing; pointing, acquisition, and tracking & CCSDS & Consultative Committee for Space Data Systems \\
CSP & Communications Services Project & DLR & German Aerospace Center \\
DoD & U.S. Department of Defense & DSOC & Deep Space Optical Communications \\
DTE & Direct to Earth & EDRS & European Data Relay System \\
EO & Earth observation & EONN & European Optical Nucleus Network \\
ESA & European Space Agency & EU & European Union \\
FEC & Forward-error correction & FSO & Free-space optical \\
GEO & Geostationary Earth orbit & GSaaS & Ground-station as a service \\
HPE & High-photon-efficiency & ILLUMA-T & Integrated LCRD Low Earth Orbit User Modem and Amplifier Terminal \\
ISR & Intelligence, surveillance, and reconnaissance & ISRO & Indian Space Research Organisation \\
ISS & International Space Station & JAXA & Japan Aerospace Exploration Agency \\
KSAT & Kongsberg Satellite Services & LCOT & Low-Cost Optical Terminal \\
LCRD & Laser Communications Relay Demonstration & LEO & Low Earth orbit \\
LUCAS & Laser Utilizing Communication System & MEO & Medium Earth orbit \\
MPLC & Multi-plane light conversion & NASA & National Aeronautics and Space Administration \\
NEC & Nippon Electric Company & NGSO & Non-geostationary orbit \\
NOC & Network operations center & O2O & Orion Artemis II Optical Communications \\
OCT & Optical communications terminal & OGS & Optical ground station \\
ONN & Optical Nucleus Network & OOK & On-off keying \\
PIN & Positive-intrinsic-negative photodiode & PPM & Pulse-position modulation \\
QKD & Quantum key distribution & RF & Radio frequency \\
SCPPM & Serially concatenated pulse-position modulation & SDA & Space Development Agency \\
SNSPD & Superconducting nanowire single-photon detector & SSC & Swedish Space Corporation \\
TBIRD & TeraByte InfraRed Delivery & TRL & Technology readiness level \\
TT\&C & Telemetry, tracking, and command & USD & U.S. dollar \\
\bottomrule
\end{tabular}
\end{table}

The technology is no longer limited to laboratory demonstrations. The National Aeronautics and Space Administration (NASA) TeraByte InfraRed Delivery (TBIRD) mission demonstrated \SI{200}{\Gbps} from low Earth orbit (LEO) and delivered \SI{4.8}{\TB} in a single pass \cite{NASA_TBIRD_2024,Wang_TBIRD_2025,Riesing_TBIRD_2025}. Airbus's SpaceDataHighway / European Data Relay System (EDRS) reported more than 80{,}000 successful laser connections over its first eight years of routine operations, together with a \SI{99.53}{\percent} reliability metric for those connections, on its October~2024 public service page \cite{Airbus_EDRS_2024}. The Japan Aerospace Exploration Agency (JAXA) and Nippon Electric Company (NEC) demonstrated a \SI{1.8}{\Gbps} relay between LEO and geostationary Earth orbit (GEO) over roughly \SI{40{,}000}{km} using the Laser Utilizing Communication System (LUCAS) \cite{JAXA_LUCAS_2025,NEC_LUCAS_2025}. NASA's Deep Space Optical Communications (DSOC) experiment demonstrated \SI{25}{\Mbps} from interplanetary range \cite{NASA_DSOC_2024}. The European Space Agency (ESA), the German Aerospace Center (DLR), Kongsberg Satellite Services (KSAT), Swedish Space Corporation (SSC), and several commercial suppliers are now building explicit OGS network services rather than one-off research facilities \cite{ESA_ONN_2026,ESA_EONN_2024,KSAT_Optical_2025,SSC_Optical_2025}. These records establish OGSs as mission-relevant infrastructure in selected segments, but they do not by themselves establish universal utility-grade service.

The remaining gap is the service layer. Optical link physics, pointing sensitivity, atmospheric turbulence, cloud outage, and RF/optical complementarity are well understood at first order. What remains less integrated in the literature is a quantitative systems view that connects these ingredients to service availability, diversity economics, standards-based interoperability, and the public industrial structure through which OGS capacity is procured. A customer does not buy peak line rate alone; it buys usable contacts, latency, service assurance, backhaul, and operational integration.

This paper therefore makes a systems-level contribution rather than a new optical-physics analysis. It combines public mission and operator records with scalar link-budget, photon-counting, availability, and cost-allocation models to identify the conditions under which OGSs become economically and operationally attractive. The central question is not whether optical links can close under controlled or bounded operational conditions; the central question is when weather-diverse, standards-compatible optical ground service becomes preferable to RF-only or hybrid RF/optical architectures.

The evidence and models are bounded explicitly. Reported rates, relay counts, and availability values are treated as demonstrated or operational only when tied to public mission or operator records. Company-primary material is used to establish product existence, advertised capability, contract announcements, or strategic positioning; it is not used by itself to establish operational leadership. Company and demand-creator comparisons are public-evidence, role-based assessments, not proprietary performance audits, investment rankings, or claims about undisclosed contracts. Appendix~\ref{app:scoring} gives the source classes, weighting rules, source inventory, and sensitivity checks. The link-budget, photon-counting, availability, and cost equations are planning and scaling models; mission-level validation would require measured pointing-acquisition-tracking covariance, site-to-site cloud correlation, background and detector calibration, modem implementation loss, safety-deconfliction, backhaul and cybersecurity analysis, and closed-loop operational statistics.

The main results are conditional. Public demonstrations now span deep-space, cislunar, relay, and high-rate LEO direct-to-ground links, so peak optical line rate is no longer the main limiting variable in every segment. Instead, the controlling variables are usable duty factor, effective site diversity, shared-network utilization, interoperability, and service-class assurance. Under the explicit normalization used later in the paper, a \SI{10}{\Gbps} near-Earth station with \$2~million/year annualized cost, 0.5~h/day of scheduled optical time, and delivered-throughput efficiency $\eta=0.7$ gives a fixed-cost component of order \$3--4k/TB. That value is a sensitivity anchor, not a tariff forecast. The robust result is that OGSs become attractive first for high-value buffered throughput and last for high-assurance continuity classes; hybrid optical+RF remains the defensible 2030 planning baseline unless anchor procurement, effective weather diversity, shared utilization, and standards maturity improve together.

The paper is organized as follows. Section~\ref{sec:technology} develops the OGS architecture and link-physics trade space, including spot size, pointing, turbulence, waveform regime, and interoperability. Section~\ref{sec:readiness} reviews public operational benchmarks and separates point-technology maturity from service-layer maturity. Section~\ref{sec:economics} develops delivered-service, availability, diversity, and shared-network cost models and identifies the boundary between OGS-only and hybrid optical+RF architectures. Section~\ref{sec:market} maps the public-evidence industrial structure by value-chain role. Section~\ref{sec:demand} assesses demand creators and scenario outcomes through 2030. Section~\ref{sec:conclusion} summarizes the conditional systems-engineering conclusion. Appendices~\ref{app:scoring}--\ref{app:cailabs} collect the public-evidence methodology, supporting cost relations, company comparison tables, and a cross-layer classification example.

\section{Technology architecture and trade space}
\label{sec:technology}

\subsection{Functional decomposition of an OGS}

An OGS is more than a telescope. It is a layered opto-electronic communications terminal with tightly coupled sensing, control, and atmospheric-interface functions. Table~\ref{tab:stack} summarizes the main subsystem stack.

Near-Earth OGS economics increasingly favor standardized, observatory-derived apertures and telecom-derived optical components. NASA's Low-Cost Optical Terminal uses a commercially available \SI{70}{cm} telescope and was explicitly conceived to reduce the cost and mission specificity of optical ground terminals \cite{NASA_LCOT_2025}. KSAT's Nemea system is a \SI{0.5}{m} class OGS designed to approach \RF-service economics \cite{KSAT_Optical_2025}. At the opposite end of the spectrum, deep-space optical support can require multi-meter-class apertures and bespoke infrastructure; NASA estimated roughly \$120~million for a \SI{12}{m} deep-space optical ground telescope concept \cite{NASA_DSOGT_Cost_2019}. This divergence between near-Earth and deep-space infrastructure economics is a central structural fact in the sector.

Deep-space ground-segment capability also broadened during 2025. ESA established its first optical communication link with NASA's DSOC experiment in July~2025 at a distance of about \SI{1.8}{AU} (\(\sim\)\SI{265}{million~km}), and in August~2025 completed a 300-million-km-class optical campaign using European ground assets. Together with NASA's September~2025 paper that DSOC exceeded all of its technical goals, these results support the assessment that deep-space optical communications are technically credible but not yet a routine shared service \cite{ESA_DSOC_Link_2025,ESA_DSOC_Campaign_2025,NASA_DSOC_Exceeds_2025}.

\begin{table}[htbp]
\centering
\caption{Functional decomposition of an OGS and the principal implementation choices that drive performance and cost.}
\label{tab:stack}
\setlength{\tabcolsep}{4pt}
\renewcommand{\arraystretch}{1.18}
\begin{tabular}{@{}p{0.11\textwidth}p{0.16\textwidth}p{0.27\textwidth}p{0.20\textwidth}p{0.15\textwidth}@{}}
\toprule
Layer & Primary function & Typical implementation choices & Principal design drivers & Representative public anchors \\
\midrule
Optical aperture and mount & Collect / transmit optical power and provide coarse line-of-sight control & \SIrange{0.4}{1.0}{m} class telescopes for near-Earth service; \SIrange{3}{12}{m} class concepts for deep-space support & Aperture cost, slew rate, enclosure design, windshake, and maintainability & LCOT, KSAT Nemea, DSOGT concepts \cite{NASA_LCOT_2025,KSAT_Optical_2025,NASA_DSOGT_Cost_2019} \\
Pointing, acquisition, and tracking (\ATP) & Establish and maintain the optical line of sight & Beacon lasers, fine-steering mirrors, tip-tilt stages, focal-plane cameras, ephemeris-aided acquisition & Pointing error budget, control bandwidth, acquisition time, body/satellite dynamics & OGS-1, LUCAS, O2O \cite{NASA_OGS1_2024,JAXA_LUCAS_2025,NASA_ArtemisII_O2O_2025} \\
Atmospheric interface & Mitigate turbulence and improve received coupling efficiency & Tip-tilt only, full \AO, single-mode fiber injection, photonic mode sorting, uplink beam combining & Fried parameter $r_0$, scintillation, daytime background, coupling stability & NASA OGS-1, TNO/FSO, Cailabs \cite{NASA_OGS1_2024,TNO_FSO_2023,Cailabs_Lasercom_2025} \\
Optical source and receiver & Generate and detect the waveform & 1.5\,$\mu$m telecom lasers, direct or coherent detection, APD/PIN chains, photon counting, SNSPD concepts & Linewidth, quantum efficiency, noise figure, eye safety, detector operating temperature & LUCAS, DSOC, RealTOR \cite{JAXA_LUCAS_2025,NASA_DSOC_2024,RealTOR_2024} \\
Modem and coding & Recover information and maintain synchronization & \CCSDS\ O3K, HPE/SCPPM, OOK, DPSK, PPM, strong FEC and interleaving & Spectral efficiency versus photon efficiency, latency, clock recovery, interoperability & CCSDS optical standards, SDA OCT \cite{CCSDS_Physical_2019,SDA_OCT_2025} \\
Safety and automation & Ensure safe laser operation and scalable remote service & Aircraft deconfliction, weather stations, cloud sensors, all-sky imagers, autonomous scheduling, NOC integration & Regulatory compliance, staffing model, false-abort rate, unattended operations & EONN, KSAT, SSC \cite{ESA_EONN_2024,KSAT_Optical_2025,SSC_Optical_2025} \\
Network interface & Deliver user data into terrestrial infrastructure & Backhaul, cloud buckets, mission ops integration, security partitioning, service telemetry & Customer abstraction, service assurance, cyber posture, scalable operations & SSC, NASA CSP, KSAT \cite{SSC_Optical_2025,NASA_CommercialPush_2025,KSAT_Optical_2025} \\
\bottomrule
\end{tabular}
\end{table}

\subsection{Link physics, spot size, and pointing sensitivity}
\label{sec:linkphysics}

\begin{figure}[htbp]
\centering
\includegraphics[width=0.52\linewidth]{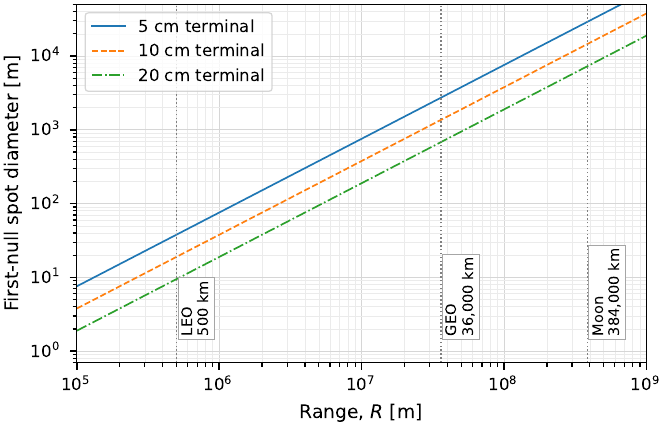}
\caption{First-null diffraction-limited spot diameter versus range for representative \SI{5}{cm}, \SI{10}{cm}, and \SI{20}{cm} transmit apertures at \SI{1{,}550}{nm}, computed from Eq.~\eqref{eq:spotdiam}. Atmospheric broadening, pointing loss, and implementation loss are excluded; the curves are geometric lower bounds on illuminated footprint.}
\label{fig:spot}
\end{figure}

For a diffraction-limited circular transmitter, the angular radius to the first Airy minimum is approximately
\begin{equation}
\theta_{\rm Airy} \approx 1.22\,\frac{\lambda}{D_t},
\label{eq:divergence}
\end{equation}
where $\lambda$ is wavelength and $D_t$ is the transmit aperture. At \SI{1{,}550}{nm}, a \SI{10}{cm} terminal gives
$\theta_{\rm Airy}\approx \SI{18.9}\,\mu{\rm rad}$ and a \SI{20}{cm} terminal gives about $\SI{9.5}\,\mu{\rm rad}$.
The corresponding first-null spot \emph{diameter} at range $R$ is
\begin{equation}
d_{\mathrm{spot}} \approx 2\,\theta_{\rm Airy}R\approx 2.44\,\frac{\lambda}{D_t}R.
\label{eq:spotdiam}
\end{equation}
If a Gaussian-equivalent $1/e^2$ spot size is preferred for detailed link-budget work, the numerical coefficient differs, but the scaling with $R\lambda/D_t$ remains unchanged. Representative first-null spot diameters for \LEO, \GEO, and lunar ranges are summarized in Table~\ref{tab:spot}, while Fig.~\ref{fig:spot} shows the same scaling continuously over range.

\begin{table}[htbp]
\centering
\caption{Representative first-null diffraction-limited spot diameters at \SI{1{,}550}{nm}. Values neglect atmospheric broadening and implementation loss.}
\label{tab:spot}
\setlength{\tabcolsep}{4pt}
\renewcommand{\arraystretch}{1.18}
\begin{tabular}{@{}p{0.17\textwidth}p{0.13\textwidth}p{0.12\textwidth}p{0.12\textwidth}p{0.38\textwidth}@{}}
\toprule
Regime & Range & $D_t=\SI{10}{cm}$ & $D_t=\SI{20}{cm}$ & Operational implication \\
\midrule
\LEO\ direct-to-ground & \SI{500}{km} & \SI{18.9}{m} & \SI{9.5}{m} & Sub-10\,$\mu$rad pointing errors already affect the power budget. \\
\GEO\ relay/feeder link & \SI{36{,}000}{km} & \SI{1.36}{km} & \SI{681}{m} & Acquisition and ephemeris quality are first-order concerns. \\
Moon-Earth & \SI{384{,}000}{km} & \SI{14.5}{km} & \SI{7.26}{km} & Photon efficiency/search-window management become dominant. \\
\bottomrule
\end{tabular}
\end{table}

A convenient Friis-like received-power expression is
\begin{equation}
P_r \approx P_t\,\eta_t\eta_r\,T_{\mathrm{atm}}\,L_p\,
\left(\frac{\pi D_t D_r}{4\lambda R}\right)^2,
\label{eq:friis}
\end{equation}
where $P_t$ is optical transmit power, $\eta_t$ and $\eta_r$ are terminal efficiencies, $T_{\mathrm{atm}}$ is atmospheric transmission, $L_p$ lumps pointing and implementation loss, and $D_r$ is the receive aperture. Eq.~(\ref{eq:friis}) makes the OGS design problem explicit: aperture, pointing, atmosphere, and receiver technology all enter multiplicatively. The clear-sky link margin can be large, while cloud cover remains a hard outage.

For the aperture and range classes considered here, the Fraunhofer far-field condition is comfortably satisfied. Even for $D_t=\SI{20}{cm}$ at $\lambda=\SI{1{,}550}{nm}$, the usual Fraunhofer distance $2D_t^2/\lambda \approx \SI{52}{km}$ is far below the \LEO, \GEO, and lunar ranges considered in Table~\ref{tab:spot}. Eq.~(\ref{eq:friis}) is therefore appropriate as a first-order comparative scaling law, provided that atmospheric loss, pointing loss, and receiver-dependent coupling are interpreted through the lumped factors $T_{\mathrm{atm}}$, $L_p$, and, if needed, an added coupling term $\eta_{\mathrm{cpl}}$.

Mispointing can consume margin quickly. A useful approximation for a near-Gaussian beam is
\begin{equation}
L_p \approx \exp\!\left[-2\left(\frac{\sigma_\theta}{\theta}\right)^2\right],
\label{eq:pointloss}
\end{equation}
where $\sigma_\theta$ is the RMS pointing error and $\theta$ is the characteristic Gaussian beam radius used in the jitter model, not automatically the Airy first-null angle. If the \SI{10}{cm}, \SI{1{,}550}{nm} first-null angle is used as an order-of-magnitude beam scale, $\sigma_\theta=\SI{5}\,\mu{\rm rad}$ gives roughly \SI{0.6}{dB} loss and $\sigma_\theta=\SI{10}\,\mu{\rm rad}$ gives roughly \SI{2.4}{dB}. These are not catastrophic values, but they are large enough to matter in acquisition design, turbulence fades, and operational margins \cite{NASA_OGS1_2024,JAXA_LUCAS_2025}.

A separate term in the pointing problem is deterministic point-ahead / velocity aberration. For a low-Earth-orbit downlink with transverse speed $v_\perp\simeq7.5\,\mathrm{km\,s^{-1}}$, the two-way point-ahead scale is $\theta_{\rm PA}\simeq2v_\perp/c\simeq50\,\mu\mathrm{rad}$. This angle is larger than the diffraction angles in Table~\ref{tab:spot}, but it is not a stochastic loss term in Eq.~\eqref{eq:pointloss}. It is a deterministic acquisition, tracking, and terminal-control requirement that must be modeled or compensated by the spacecraft/ground control loop.

\begin{figure}[htbp]
\centering
\includegraphics[width=0.52\linewidth]{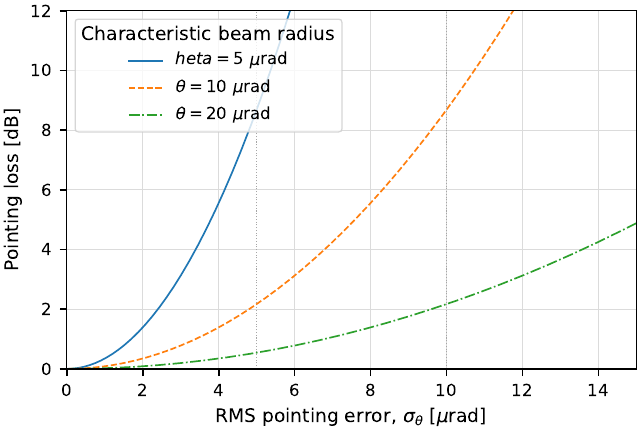}
\caption{Approximate pointing-loss sensitivity for the near-Gaussian model in Eq.~\eqref{eq:pointloss}. The curves use characteristic beam radii of 5~\(\mu\)rad, 10~\(\mu\)rad, and 20~\(\mu\)rad; the calculation is a scalar dB-sensitivity model and is not a substitute for a mission-specific pointing-acquisition-tracking covariance budget.}
\label{fig:pointloss}
\end{figure}

Eqs.~(\ref{eq:friis}) and (\ref{eq:pointloss}) are intentionally first-order. Eq.~(\ref{eq:friis}) assumes ideal circular apertures and aggregates implementation-specific penalties into scalar loss terms; it is therefore suitable for architecture comparison, not for final end-to-end receiver design. Eq.~(\ref{eq:pointloss}) assumes a near-Gaussian far-field profile and stochastic pointing jitter, and should be interpreted as an approximate dB-sensitivity model rather than as a substitute for full PAT covariance analysis. Figure~\ref{fig:pointloss} plots this scalar sensitivity for characteristic beam radii representative of narrow optical links.

\subsection{Technology families by mission class}

The phrase ``optical communications'' hides several materially different technical regimes. Near-Earth, high-photon-flux links can use telecom-derived components, preamplified coherent or quasi-coherent receivers, and formats optimized for spectral efficiency. Cislunar and deep-space links operate closer to the photon-starved regime and prioritize photons per bit rather than bits per hertz. Relay-centric architectures shift the geometry burden into space; direct-to-ground architectures put more stress on weather diversity and pass scheduling. Table~\ref{tab:families} organizes this trade space.

\begin{table}[htbp]
\centering
\caption{Technology-family comparison across major OGS-relevant mission classes. Values are representative, not prescriptive, and are intended to distinguish the dominant engineering regime in each segment.}
\label{tab:families}
\setlength{\tabcolsep}{3pt}
\renewcommand{\arraystretch}{1.18}
\begin{tabular}{@{}p{0.12\textwidth}p{0.10\textwidth}p{0.20\textwidth}p{0.09\textwidth}p{0.09\textwidth}p{0.17\textwidth}p{0.15\textwidth}@{}}
\toprule
Mission class & Typical range & Waveform / receiver regime & Space aperture & OGS aperture & Ground challenge & Public anchors \\
\midrule
High-rate \LEO\ direct-to-ground & \SIrange{500}{1200}{km} slant range & High-photon-flux near-infrared links; O3K-class or similar; direct or coherent-style reception & \SIrange{5}{15}{cm} & \SIrange{0.4}{1.0}{m} & Short contact windows, turbulence-induced coupling loss, scheduling efficiency & TBIRD, AIRSAT-02, LCOT \cite{NASA_TBIRD_2024,CAS_AIRSAT02_2026,NASA_LCOT_2025} \\
\GEO\ optical relay & \SI{36{,}000}{km} to\newline \SI{45{,}000}{km} & Telecom-style \SI{1{,}550}{nm} relay links; high pointing precision; managed geometry & \SIrange{8}{15}{cm} relay\newline terminals & \SIrange{0.5}{1.0}{m} plus engineered diversity & Acquisition reliability, service continuity, network operations & EDRS, LUCAS, LCRD / ILLUMA-T \cite{Airbus_EDRS_2024,JAXA_LUCAS_2025,NASA_ILLUMAT_2025} \\
Cislunar \DTE/relay & \SI{380{,}000}{km} to\newline \SI{450{,}000}{km} & Photon-efficient PPM / HPE-style coding; strong timing and synchronization discipline & \SIrange{8}{20}{cm} & \SIrange{0.6}{1.0}{m} and sometimes larger & Background rejection, cloud diversity, mixed service requirements & O2O, Moonlight \cite{NASA_O2O_2024,NASA_ArtemisII_O2O_2025,ESA_Moonlight_2024} \\
Deep-space direct-to-ground & Inter\-planetary & Photon-starved regime; strong coding and often photon counting or very low-noise reception & Mission specific & \SIrange{3}{12}{m} class concepts & Aperture cost, link margin, low network utilization & DSOC, RealTOR, DSOGT concepts \cite{NASA_DSOC_2024,RealTOR_2024,NASA_DSOGT_Cost_2019} \\
Telecom feeder links & \GEO\ or MEO gateway & Very-high-throughput, high-availability links; ultimately toward coherent and Tb/s scaling & Multi-beam feeder\newline terminals & Multi-site gateway network & Carrier-grade availability, uplink turbulence mitigation, cloud diversity & ESA feeder-link work, HydRON, SES testing \cite{ESA_FeederLink_2021,ESA_HydRON_2025,SES_Cailabs_2025} \\
\bottomrule
\end{tabular}
\end{table}

A useful quantitative discriminator is the received photons per information bit,
\begin{equation}
N_b \approx \frac{P_r\lambda}{hc\,R_b},
\label{eq:photonsperbit}
\end{equation}
where $P_r$ is received optical power and $R_b$ is information rate. Equation~\eqref{eq:photonsperbit} is a signal-only photon-count estimate: it omits background photons, detector dark counts, excess noise, implementation loss, and coding margins, and therefore represents an upper bound on usable signal photons rather than a complete link margin. In this section $R_b$ is the delivered information rate; when a raw physical line rate is quoted instead, FEC, framing, and protocol overhead must be absorbed into an effective information-rate conversion or into the efficiency factor used in the cost model. When $N_b$ is large, spectral efficiency and hardware simplicity dominate; when $N_b$ is small, photon efficiency dominates, pushing the system toward PPM-like formats, stronger coding, and very low-noise or photon-counting receivers. This is why it is technically inaccurate to treat \LEO\ bulk downlink, lunar relay, and deep-space optical communications as though they were one homogeneous technology stack.

For illustration, $P_r=\SI{1}{nW}$ at $\lambda=\SI{1{,}550}{nm}$ yields $N_b \approx 7.8$ photons/bit at $R_b=\SI{1}{Gbps}$, but $N_b \approx 78$ photons/bit at $R_b=\SI{100}{Mbps}$. This simple scaling explains why otherwise similar links can fall into very different waveform regimes as received power and target rate change.

A compact radiometric check gives the same regime separation. With $P_t=\SI{1}{W}$, $D_t=\SI{10}{cm}$, $D_r=\SI{0.5}{m}$, $\lambda=\SI{1{,}550}{nm}$, and an aggregate factor $\eta_t\eta_rT_{\mathrm{atm}}L_p=0.1$, Eq.~\eqref{eq:friis} gives received powers of order $2.6\times10^{-4}$~W at \SI{500}{km}, $5\times10^{-8}$~W at \GEO\ range, and $4\times10^{-10}$~W at lunar range. The exact values are implementation dependent, but the $R^{-2}$ collapse explains why near-Earth, relay, and cislunar/deep-space OGSs cannot be validated with one scalar performance number.

\subsection{Waveform, modulation, and receiver regimes}

Optical-space-communication systems are now clearly splitting into at least two operational waveform regimes. The first is the \emph{high-photon-flux near-Earth regime}, where designers can exploit telecom-derived lasers, preamplified coherent or near-coherent receivers, narrow-linewidth local oscillators, and formats optimized for rate and implementation efficiency. The second is the \emph{photon-efficient regime}, where received photons per information bit are scarce enough that PPM-like formats, strong coding, precise synchronization, and photon-counting or extremely low-noise receivers dominate the design. The public system map reflects this split. LCRD and ILLUMA-T are near-Earth relay systems built around a telecom-style \SI{1{,}550}{nm} ecosystem \cite{NASA_ILLUMAT_2025,NASA_OGS1_2024}. O2O uses \CCSDS-standard PPM for cislunar operations \cite{NASA_ArtemisII_O2O_2025}. DSOC and RealTOR represent the deep-space / photon-efficient end of the market, where receiver noise, timing recovery, and coding gain become as important as aperture size \cite{NASA_DSOC_2024,RealTOR_2024,CCSDS_Physical_2019}.

A practical implication for OGS design is that the receive chain is not a commodity element across all market segments. A station optimized for \SI{10}{\Gbps}-class near-Earth service may prioritize single-mode coupling efficiency, low-jitter tracking, and compatibility with coherent or preamplified photonic hardware. A cislunar or deep-space terminal, by contrast, may accept much lower raw symbol rates but demand tighter timing distribution, narrower optical filtering, different detector technology, and stronger coding/interleaving. This is one reason the optical-ground market should not be analyzed as a monolith: the supply chain overlaps, but the performance bottlenecks differ sharply by mission class.

\subsection{Atmospheric channel and mitigation trade space}

The Earth atmosphere is the defining challenge of the OGS. Clouds are a hard outage. Turbulence, scintillation, beam wander, and daytime background are softer but still mission-defining penalties. A common scalar measure of turbulence severity is the ratio $D/r_0$, where $D$ is aperture diameter and $r_0$ is the Fried coherence length. For a representative \SI{0.5}{m} OGS aperture and the LCRD site data used in NASA availability work, the turbulence-severity parameter is approximately $D/r_0\approx2.50$ at Table Mountain, $2.19$ at White Sands, and $1.15$ at Haleakala, which quantitatively explains why site selection materially affects fiber-coupling and high-order correction difficulty \cite{LCRD_Availability_2017}. A site with twice the effective $r_0$ is not marginally better; it can be in a meaningfully different operational regime.

Table~\ref{tab:rzero} translates the same public $r_0$ values into explicit aperture-normalized difficulty. The message is practical rather than abstract: a \SI{0.5}{m} station at Haleakala operates close to the transition where high-coupling-efficiency reception is comparatively tractable, while a \SI{1.0}{m} station at Table Mountain is already in a regime where high-order correction or a more tolerant receive architecture becomes materially more important. Site selection is therefore not merely a question of cloud fraction; it is also a question of how expensive it will be to realize the desired coupling efficiency and margin.

Daytime and low-elevation operation add receiver-background and calibration constraints that should not be hidden in a single efficiency factor. Solar and lunar radiance, detector dark current, narrowband-filter leakage, coupling fluctuations, cloud-screening thresholds, and aircraft-safety abort logic all enter the usable-link probability. In this paper those effects are aggregated into $T_{\mathrm{atm}}$, $L_p$, $\eta_{\mathrm{cpl}}$, and the delivered-throughput factor $\eta$; a flight service would require separate calibration of each contribution.

\begin{table}[htbp]
\centering
\caption{Illustrative turbulence severity from public LCRD median \SI{1{,}550}{nm} Fried-parameter values \cite{LCRD_Availability_2017}. The ratio $D/r_0$ is a compact proxy for wavefront-correction difficulty.}
\label{tab:rzero}
\setlength{\tabcolsep}{4pt}
\renewcommand{\arraystretch}{1.18}
\begin{tabular}{@{}p{0.14\textwidth}p{0.14\textwidth}p{0.10\textwidth}p{0.10\textwidth}p{0.38\textwidth}@{}}
\toprule
Site & Median $r_0$ & $D/r_0$ for & $D/r_0$ for & Practical implication \\
 & at \SI{1{,}550}{nm} & \SI{0.5}{m} OGS & \ \SI{1.0}{m} OGS &  \\
\midrule
Table Mountain & \SI{20.0}{cm} & 2.50 & 5.00 & Strong turbulence burden; coupling efficiency and uplink pre-compensation become more demanding. \\
White Sands & \SI{22.8}{cm} & 2.19 & 4.39 & Moderately demanding; operationally useful but still clearly \AO-relevant. \\
Haleakala & \SI{43.5}{cm} & 1.15 & 2.30 & Most benign of the three; particularly attractive for high-coupling-efficiency reception. \\
\bottomrule
\end{tabular}
\end{table}

The canonical turbulence scalings are the Fried parameter $r_0$, the coherence time $\tau_0\simeq0.31 r_0/v_{\rm eff}$, and the Greenwood frequency $f_G\simeq0.43 v_{\rm eff}/r_0$ for effective transverse wind speed $v_{\rm eff}$ \cite{Fried1966,AndrewsPhillips2005,Greenwood1977}. For $r_0=\SI{20}{cm}$ and $v_{\rm eff}=30$--$50\,\mathrm{m\,s^{-1}}$, this gives $f_G\sim\SIrange{65}{110}{Hz}$, so closed-loop adaptive-optics bandwidths of several hundred hertz are a plausible engineering scale rather than a free parameter.

Telecom feeder links show how demanding the atmospheric problem becomes once availability requirements rise. ESA's feeder-link work states explicitly that cloud mitigation may require on the order of \SI{10} gateway sites separated by at least about \SI{500}{km} to reduce weather correlation, and that uplink pre-compensation may require adaptive-optics bandwidth exceeding roughly \SI{300}{Hz} \cite{ESA_FeederLink_2021}. These numbers are strategically important because they show that the challenge is not simply to ``build a larger telescope.'' It is to field a geographically distributed, fast-control, carrier-grade optical network.

Three technical philosophies currently dominate atmospheric mitigation:
\begin{enumerate}
\item classical \AO\ and fast steering, which seek to correct the wavefront directly and maximize coupling into a small field of view or a single-mode fiber;
\item photonic mode sorting / beam combining, exemplified by MPLC-based approaches that improve robustness without relying exclusively on conventional deformable-mirror architectures; and
\item coding/interleaving hardening, which treats some atmospheric impairment as a burst-error problem and recovers performance at the modem level.
\end{enumerate}
In practice, high-performance OGSs increasingly combine all three. Table~\ref{tab:mitigation} compares the approaches.

\begin{table}[htbp]
\centering
\caption{Comparison of the main atmospheric-mitigation strategies now visible in the OGS market.}
\label{tab:mitigation}
\setlength{\tabcolsep}{4pt}
\renewcommand{\arraystretch}{1.18}
\begin{tabular}{@{}p{0.13\textwidth}p{0.18\textwidth}p{0.20\textwidth}p{0.18\textwidth}p{0.24\textwidth}@{}}
\toprule
Approach & Physical principle & Best-fit use cases & Strengths & Limitations / cost drivers \\
\midrule
Site diversity & Avoid clouds statistically by using geographically separated sites & Any buffered service; essential precursor to commercial OGS networks & Most powerful mitigation of cloud outage; scalable at network level & Requires multi-site capex, backhaul, scheduling, and low site-to-site weather correlation \cite{LCRD_Availability_2017,DLR_AOGSN_2025} \\
Tip-tilt and fast steering & Correct beam wander and low-order pointing disturbance & Moderate-rate \LEO\ links, acquisition support, direct-detection systems & Simpler than full \AO; high operational robustness & Cannot correct higher-order aberration or maximize single-mode coupling \cite{NASA_OGS1_2024} \\
Full adaptive optics + fiber coupling & Wavefront sensing and deformable correction prior to narrow-field or fiber receive chain & High-rate near-Earth links and coherent / single-mode architectures & Highest coupling efficiency when the atmosphere is tractable; strong path to coherent-grade links & More complex optics, calibration, maintenance, and control bandwidth requirements \cite{NASA_OGS1_2024,TNO_FSO_2023} \\
MPLC mode sorting and beam combining & Passive photonic handling of distorted spatial modes on Rx and/or diversity on Tx & 10+~\Gbps class links, feeder-link pathway, multi-mission commercial stations & Potential robustness and scalability; may reduce dependence on classical \AO-only solutions & Product maturity and ecosystem breadth still smaller than classical \AO\ supply chain \cite{Cailabs_Lasercom_2025,Cailabs_FAQ_2025} \\
Coding / interleaving hardening & Absorb atmospheric fades as burst errors at the modem and FEC layer & Photon-efficient cislunar / deep-space links and robust DTE links & Powerful when photons are scarce or fades are bursty; interoperable through standards & Does not solve cloud outage; cannot recover power-budget deficits that are too severe \cite{CCSDS_Physical_2019,NASA_DSOC_2024} \\
\bottomrule
\end{tabular}
\end{table}

\subsection{Standards and interoperability}

The OGS market cannot scale without interoperability. On the civil side, the \CCSDS\ optical standards now define physical-layer and coding/synchronization frameworks for both high-photon-flux and high-photon-efficiency regimes \cite{CCSDS_Physical_2019}. On the defense side, the \SDA\ OCT standards are shaping terminal and gateway design for proliferated optical networking \cite{SDA_OCT_2025}. The commercial significance is substantial. In a proprietary world, every spacecraft terminal and every OGS network must be co-engineered. In a standards-based world, OGS time can be sold as a service. ESA's ONN/EONN effort states explicitly that the space segment is increasingly ready while the limiting factor is the lack of a reliable operational ground segment and the need for consistent interfaces \cite{ESA_ONN_2026,ESA_EONN_2024}. Standards convergence is therefore not peripheral policy work; it is the precondition for an investable OGS service layer.

\section{Global state of the art and readiness}
\label{sec:readiness}

\subsection{Operational benchmarks and demonstrations}

\begin{figure}[htbp]
\centering
\includegraphics[width=0.52\linewidth]{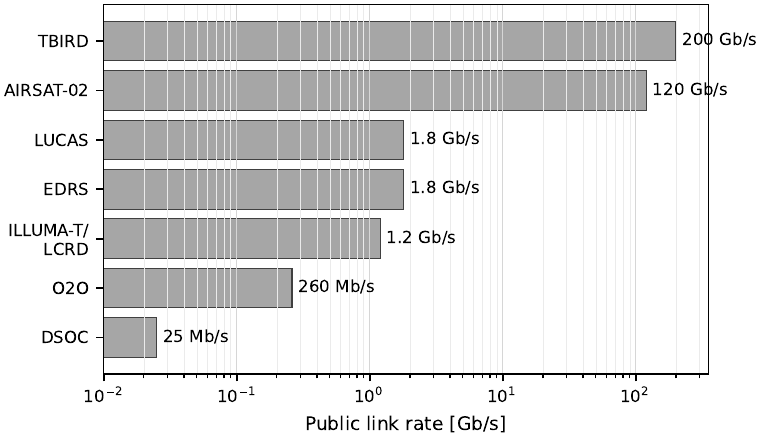}
\caption{Selected public optical-communication rate benchmarks used in Table~\ref{tab:demos}. Values are plotted on a logarithmic axis to compare mission classes. The figure is a rate comparison only; it does not measure service availability, network maturity, or commercial readiness.}
\label{fig:rates}
\end{figure}

Table~\ref{tab:demos} summarizes the public systems that matter most for judging the current state of the art. The important point is not just the spread in line rates; it is the breadth of mission classes that now have credible optical evidence. Figure~\ref{fig:rates} plots the same public rate benchmarks on a logarithmic scale to emphasize the spread across mission classes without implying service readiness.

\begin{table}[htbp]
\centering
\caption{Selected public demonstrations and operational systems relevant to OGS assessment. Rates are the best public values for the cited configuration.}
\label{tab:demos}
\setlength{\tabcolsep}{4pt}
\renewcommand{\arraystretch}{1.18}
\begin{tabular}{@{}p{0.16\textwidth}p{0.12\textwidth}p{0.12\textwidth}p{0.14\textwidth}p{0.30\textwidth}p{0.06\textwidth}@{}}
\toprule
System & Regime & Public rate & Status & Why it matters & Source \\
\midrule
TBIRD (NASA / MIT Lincoln Laboratory) & \LEO\ direct-to-ground & \SI{200}{\Gbps} & Demo.\newline completed & High-rate \LEO\ downlink benchmark; \SI{4.8}{\TB} in one pass & \cite{NASA_TBIRD_2024,Wang_TBIRD_2025,Riesing_TBIRD_2025} \\
ILLUMA-T + LCRD & ISS relay via \GEO & \SI{1.2}{\Gbps} class & Operational experiment completed & End-to-end relay architecture with flight operations, ground procedures, and dual OGS support & \cite{NASA_ILLUMAT_2025,NASA_OGS1_2024} \\
EDRS / SpaceData-Highway & \GEO\ relay service & up to \SI{1.8}{\Gbps} & Operational & Documented commercial relay-service benchmark; Airbus's October~2024 page reports 80{,}000+ laser links and \SI{99.53}{\percent} reliability for that service metric & \cite{Airbus_EDRS_2024} \\
LUCAS (JAXA / NEC) & \LEO-\GEO\ relay & \SI{1.8}{\Gbps} & Operational\newline demo. & Sovereign Japanese relay capability at \SI{1.5}{\um} and \SI{40{,}000}{km} separation & \cite{JAXA_LUCAS_2025,NEC_LUCAS_2025} \\
O2O / Artemis II & Moon-Earth & up to \SI{260}{\Mbps} down / \SI{20}{\Mbps} up & Flight demonstration path & Anchors cislunar optical service economics and requirements & \cite{NASA_O2O_2024,NASA_ArtemisII_O2O_2025} \\
DSOC & Deep-space direct-to-ground & \SI{25}{\Mbps} & Demo.\newline completed & Public deep-space optical benchmark; validates photon-efficient regime & \cite{NASA_DSOC_2024} \\
AIRSAT-02 / Pamir OGS & \LEO\ direct-to-ground & \SI{120}{\Gbps} peak & Operational\newline demo. & High-altitude sovereign OGS infrastructure and high-rate DTE claims in China & \cite{CAS_Pamir_2024,CAS_AIRSAT02_2026} \\
\bottomrule
\end{tabular}
\end{table}

\subsection{Readiness by segment: point technology versus service layer}

\begin{figure}[htbp]
\centering
\includegraphics[width=0.52\linewidth]{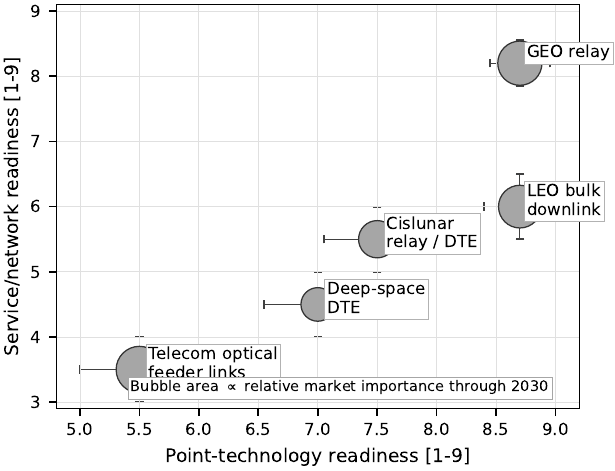}
\caption{Ordinal readiness map by segment. Symbols mark the midpoints of the author-synthesis ranges in Table~\ref{tab:readiness}; horizontal and vertical error bars show the range extents. Bubble area indicates qualitative market relevance through 2030. The axes are ordinal public-evidence assessments, not measured TRL values or forecasts.}
\label{fig:readiness}
\end{figure}

The correct readiness framework has two axes: point technology and service layer. Figure~\ref{fig:readiness} visualizes the same assessment using midpoint values and range bars derived from Table~\ref{tab:readiness}. The highest-scoring point technology today is \GEO\ optical relay, while the most commercially promising but still service-constrained segment is high-rate \LEO\ direct downlink.

\begin{table}[htbp]
\centering
\caption{Comparative readiness assessment by segment. The first score evaluates whether the link technology itself is mature; the second evaluates whether the service layer is mature enough to be bought as infrastructure.}
\label{tab:readiness}
\setlength{\tabcolsep}{4pt}
\renewcommand{\arraystretch}{1.18}
\begin{tabular}{@{}p{0.16\textwidth}p{0.11\textwidth}p{0.11\textwidth}p{0.24\textwidth}p{0.30\textwidth}@{}}
\toprule
Segment & Point tech. & Service / network & Primary public anchor & Dominant blocker to broad adoption \\
\midrule
\LEO\ bulk downlink & 8.5--9.0 & 5.5--6.5 & TBIRD, AIRSAT-02, ILLUMA-T \cite{NASA_TBIRD_2024,Wang_TBIRD_2025,Riesing_TBIRD_2025,CAS_AIRSAT02_2026,NASA_ILLUMAT_2025} & Weather-diverse OGS capacity, automated booking, and network interoperability \\
\GEO\ optical relay & 8.5--9.0 & 8.0--8.5 & EDRS and LUCAS \cite{Airbus_EDRS_2024,JAXA_LUCAS_2025} & Expansion beyond mission-specific and sovereign ecosystems \\
Cislunar relay / \DTE & 7.0--8.0 & 5.0--6.0 & O2O and Moonlight demand pull \cite{NASA_O2O_2024,ESA_Moonlight_2024} & Sparse infrastructure and mixed assurance requirements \\
Deep-space \DTE & 6.5--7.5 & 4.0--5.0 & DSOC \cite{NASA_DSOC_2024} & Large-aperture economics and weak shared-network utilization \\
Telecom optical feeder links & 5.0--6.0 & 3.0--4.0 & ESA feeder-link work, HydRON, SES testing \cite{ESA_FeederLink_2021,ESA_HydRON_2025,SES_Cailabs_2025} & Carrier-grade availability through atmosphere and gateway-network economics \\
\bottomrule
\end{tabular}
\end{table}

\subsection{Availability engineering and site diversity}
\label{sec:availability}

\begin{figure}[htbp]
\centering
\includegraphics[width=0.52\linewidth]{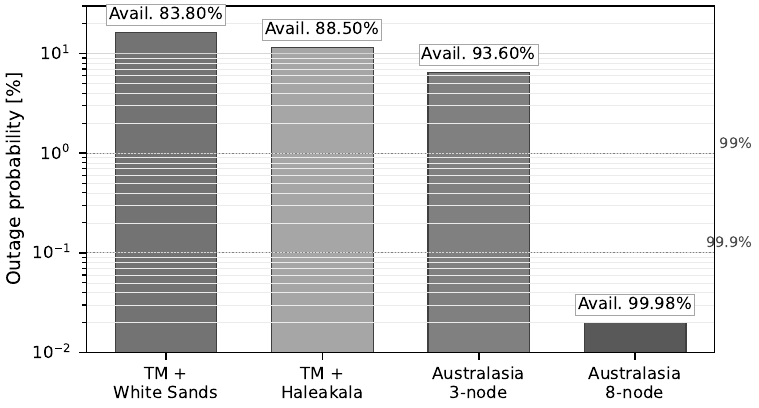}
\caption{Illustrative public availability examples expressed as outage probability. The first two bars are NASA LCRD two-site cases; the third and fourth are DLR Australasian network examples. Horizontal guides mark 99\% and 99.9\% availability thresholds. The DLR values are used as reported availability benchmarks and as the calibration cases in Table~\ref{tab:neff_anchor}; they are not a fitted weather-correlation model for Eq.~\eqref{eq:avail}.}
\label{fig:availability}
\end{figure}

Availability is the variable that most sharply separates a successful optical experiment from a communications service. NASA's LCRD site-diversity study found that Table Mountain plus White Sands yielded about \SI{83.8}{\percent} availability, while Table Mountain plus Haleakala yielded about \SI{88.5}{\percent} \cite{LCRD_Availability_2017}. DLR's Australasian network work is even more revealing from a network-design perspective: a three-node network was reported at \SI{6.4}{\percent} outage probability, while an eight-node network fell to about \SI{0.02}{\percent} outage probability \cite{DLR_AOGSN_2025}. These are large service-level differences. They show that availability is a network-size, weather-correlation, and scheduling problem, not only a telescope-quality problem. The LCRD and DLR values are load-bearing single-source public anchors in the evidence hierarchy of Appendix~\ref{app:scoring}; they are useful for calibration and sensitivity analysis, but they should not be interpreted as independent ensemble statistics. Figure~\ref{fig:availability} expresses these public cases as outage probability so that the difference between 99\% and 99.9\% service levels is visually explicit.

A useful upper-envelope approximation for statistically independent sites is
\begin{equation}
A_{\mathrm{ind}} \approx 1-\prod_{i=1}^{N_{\mathrm{phys}}}(1-A_i),
\label{eq:avail}
\end{equation}
where $N_{\mathrm{phys}}$ is the physical number of stations and $A_i$ is the availability of station $i$. Eq.~\eqref{eq:avail} is an independence bound, not a weather-correlation model. Real OGS networks can violate the assumptions through regional cloud systems, maintenance clustering, pass-geometry correlations, staffing or software common modes, laser-safety interruptions, and shared backhaul or cybersecurity constraints.

For identical sites with single-site availability $a$, a compact sensitivity parameterization is
\begin{equation}
\Avail(N_{\mathrm{phys}},a,\rho_{\mathrm{eff}})
\simeq
1-(1-a)^{N_{\mathrm{eff}}},
\qquad
N_{\mathrm{eff}}\simeq
\frac{N_{\mathrm{phys}}}{1+(N_{\mathrm{phys}}-1)\rho_{\mathrm{eff}}},
\label{eq:neff_corr}
\end{equation}
where $0\le\rho_{\mathrm{eff}}<1$ is an effective common-mode dependence parameter. It aggregates weather, maintenance, pass geometry, scheduling, and operations rather than representing a single meteorological correlation coefficient. In the absence of measured joint-weather and joint-operations statistics, $N_{\mathrm{eff}}$ should be treated as a sensitivity parameter, not as an implied prediction.

A useful calibration exercise is possible when a reported network availability is available. If homogeneous site availability $a$ is assumed, the implied effective independent-site count is
\begin{equation}
N_{\mathrm{eff}}^{\mathrm{obs}}
=
\frac{\ln(1-A_{\mathrm{net}}^{\mathrm{obs}})}{\ln(1-a)} .
\label{eq:neff_obs}
\end{equation}
Table~\ref{tab:neff_anchor} applies Eq.~\eqref{eq:neff_obs} to the DLR Australasian three-node and eight-node outage cases. The exercise is not a fitted weather-correlation model: the underlying individual-site availabilities, seasonal weights, pass geometry, and operations policy are not reconstructed here. It provides a quantitative reference point for the central uncertainty. Depending on the assumed single-site availability, the DLR eight-node result can correspond to approximately four to eight effective independent sites; the upper end occurs when individual sites are assumed less available, because more effective independent opportunities are then required to reproduce the same reported network availability. Thus the distinction between $N_{\mathrm{phys}}=8$ and $N_{\mathrm{eff}}\approx4$--8 is not a detail; it is a first-order driver of service economics.

\begin{table}[htbp]
\centering
\caption{Illustrative back-calculation of effective independent-site count from reported DLR Australasian outage examples using Eq.~\eqref{eq:neff_obs}. Values depend on the assumed homogeneous single-site availability $a$ and are used only as calibration anchors, not as fitted weather-correlation parameters.}
\label{tab:neff_anchor}
\setlength{\tabcolsep}{5pt}
\renewcommand{\arraystretch}{1.15}
\begin{tabular}{@{}p{0.28\textwidth}p{0.16\textwidth}p{0.14\textwidth}p{0.14\textwidth}p{0.14\textwidth}@{}}
\toprule
Reported case & Network availability & $N_{\mathrm{eff}}$ if $a=0.65$ & $N_{\mathrm{eff}}$ if $a=0.75$ & $N_{\mathrm{eff}}$ if $a=0.85$ \\
\midrule
Australasia 3-node & 93.6\% & 2.6 & 2.0 & 1.4 \\
Australasia 8-node & 99.98\% & $\approx8$ & 6.1 & 4.5 \\
\bottomrule
\end{tabular}
\end{table}

Annual network availability is not, by itself, a sufficient service metric for latency-critical optical communications. Let \(T_{\max}\) denote the maximum tolerable latency for successful delivery, let \(\Delta t_{\rm opp}\) denote the mean interval between useful optical opportunities, and let
\begin{equation}
K(T_{\max}) \equiv 1 + \left\lfloor \frac{T_{\max}}{\Delta t_{\rm opp}} \right\rfloor
\label{eq:kopp}
\end{equation}
be the number of opportunities available within the allowed decision window. If \(p_{{\rm pass},i,j}\) denotes the probability that station \(i\) successfully acquires, closes, and completes opportunity \(j\), then the probability of at least one successful optical service opportunity within \(T_{\max}\) is
\begin{equation}
p_{\rm svc}(T_{\max}) \approx 1 - \prod_{j=1}^{K(T_{\max})}\prod_{i=1}^{N_{\mathrm{phys}}}\Bigl(1-p_{{\rm pass},i,j}\Bigr),
\label{eq:psvc}
\end{equation}
which reduces to \(1-(1-p_{\rm pass})^{N_{\mathrm{eff}}K}\) for identical statistically independent effective opportunities. Eq.~(\ref{eq:psvc}) is optimistic if the physical site count is used in place of \(N_{\mathrm{eff}}\), because weather and operational failures are correlated. Its purpose is therefore not to predict flight performance exactly, but to make clear that latency-critical traffic is governed by decision-window success rather than annualized \(\Avail\).

The practical consequence is immediate. If \(T_{\max}<\Delta t_{\rm opp}\), so that \(K=1\), then permanent spatial diversity is often the dominant lever for service quality. If, by contrast, \(T_{\max}\gg\Delta t_{\rm opp}\) and the storage condition in Eq.~(\ref{eq:buffer}) is satisfied, buffering can substitute for some diversity. This is why a bursty \LEO\ EO mission and a defense or command-assurance service can have very different architectural implications even if their annual data volumes are similar.

\subsection{When is OGS-only rational, and when is hybrid required?}

The engineering boundary between optical-only and hybrid optical+\RF\ architectures can be written more explicitly than is usually done in market commentary. Let $\dot D$ be the average payload data-generation rate, let $B_{\mathrm{store}}$ be onboard storage, and let $\tau_{\mathrm{out}}(q)$ be the $q$-quantile of cumulative optical outage over the planning horizon relevant to operations. A necessary condition for an optical-only payload path is
\begin{equation}
B_{\mathrm{store}} \gtrsim \dot D\,\tau_{\mathrm{out}}(q),
\label{eq:buffer}
\end{equation}
with additional margin for maintenance, deferred contacts, and burst traffic. Eq.~(\ref{eq:buffer}) explains why OGS-first payload return can be defensible for buffered Earth-observation and science missions, but remains risky for service classes that cannot tolerate weather-driven delivery slips.

A second condition is operational rather than informational: command and contingency traffic must tolerate the same outage statistics as the payload path. In practice, this is where most architectures revert to hybrid. If the mission requires near-continuous TT\&C, human-rated or safety-critical operations, or a very high assured-contact probability at specific times, the incremental cost of optical site diversity is usually harder to justify than keeping an \RF\ continuity layer. This is why current \GEO\ relay, cislunar, and defense architectures are converging not to optical-only systems but to \emph{functional separation}: optical for throughput, \RF\ for assured continuity. The real design question is therefore not ``optical or RF,'' but ``which traffic classes should move to optical first, and which traffic classes still justify \RF\ fallback?'' Table~\ref{tab:hybrid_boundary} summarizes this optical-only versus hybrid boundary by mission class.
Table~\ref{tab:availability_criticality} further clarifies that annual \(\Avail\) is the relevant figure of merit for buffered traffic, whereas latency-critical service quality is better described by the decision-window success probability \(p_{\rm svc}(T_{\max})\) in Eq.~(\ref{eq:psvc}).

\begin{table}[htbp]
\centering
\caption{Decision boundary between OGS-only and hybrid architectures by mission class.}
\label{tab:hybrid_boundary}
\setlength{\tabcolsep}{4pt}
\renewcommand{\arraystretch}{1.18}
\begin{tabular}{@{}p{0.12\textwidth}p{0.10\textwidth}p{0.13\textwidth}p{0.23\textwidth}p{0.34\textwidth}@{}}
\toprule
Mission class & Storage tolerance & Continuity requirement & Technically preferred near-term architecture & Why \\
\midrule
Buffered \LEO\ EO / science & High & Moderate & OGS-first payload return with \RF\ TT\&C backup & Eq.~(\ref{eq:buffer}) is easy to satisfy with onboard storage and delayed delivery tolerance. \\[2pt]
Defense bulk ISR / mesh egress & Moderate to high & High for selected traffic classes & Hybrid optical ground entry & Optical handles bulk throughput; continuity and mission-critical flows still justify \RF\ fallback. \\[2pt]
\GEO\ relay services & Moderate & High & Engineered optical relay plus hybrid ground assurance where needed & Availability targets are high, but traffic is premium and manageable. \\[2pt]
Crewed cislunar / lunar operations & Low for critical traffic & Very high & Hybrid strongly preferred & Safety, command assurance, and schedule-critical contacts dominate. \\[2pt]
Deep-space science & High for bulk science & High mission assurance & Optical augmentation of deep-space \RF & Science return benefits from optical; mission assurance still favors a parallel \RF\ layer. \\[2pt]
Telecom feeder links & Low & Carrier-grade & Hybrid or RF-dominant until very large OGS networks exist & Carrier-grade availability through atmosphere is the central unsolved cost problem. \\
\bottomrule
\end{tabular}
\end{table}

\begin{table}[htbp]
\centering
\caption{Interpretation of annual availability by service criticality. Annual unavailability is \(8760(1-\Avail)\) h/yr. The table illustrates why annual \(\Avail\) alone is insufficient for latency-critical missions.}
\label{tab:availability_criticality}
\setlength{\tabcolsep}{4pt}
\renewcommand{\arraystretch}{1.18}
\begin{tabular}{@{}p{0.22\textwidth}p{0.10\textwidth}p{0.12\textwidth}p{0.22\textwidth}p{0.25\textwidth}@{}}
\toprule
Service class & Illustrative & Annual & More relevant metric & Architectural implication
\\
 & \(\Avail\) target & unavailability &  & 
\\
\midrule
Buffered EO / science return & 0.99 & 87.6 h/yr & Eq.~(\ref{eq:buffer}) plus revisit cadence & Next-pass recovery may be acceptable if latency is non-critical \\
Managed relay service & 0.999 & 8.76 h/yr & Annual \(\Avail\) plus scheduled-pass success & Engineered diversity desirable, but limited outages may be tolerable \\
Latency-critical mission traffic & 0.9997 & 2.63 h/yr & \(p_{\rm svc}(T_{\max})\) from Eq.~(\ref{eq:psvc}) & Annual availability is insufficient; decision-window success drives design \\
Near-continuity / command assurance & 0.9999 & 52.6 min/yr & \(p_{\rm svc}(T_{\max})\) with persistent diversity & Hybrid optical+\RF\ usually required \\
Ultra-high-assurance traffic & 0.99999 & 5.26 min/yr & Window-based success under correlated-failure assumptions & Permanent diversity and \RF\ continuity are generally unavoidable \\
\bottomrule
\end{tabular}
\end{table}

\section{Economics, scaling, and service regimes}
\label{sec:economics}

\subsection{Cost structure and delivered-service model}

Public optical-ground pricing remains sparse, so the right way to reason about economics is at the service-network level. A generic annualized network cost can be written as
\begin{equation}
\Cnet \approx \sum_{i=1}^{N}\left(C^{(i)}_{\mathrm{capex}}\cdot\mathrm{CRF}+C^{(i)}_{\mathrm{opex}}\right)
+ C_{\mathrm{NOC}}+C_{\mathrm{backhaul}}+C_{\mathrm{compliance}},
\label{eq:cnet}
\end{equation}
where the terms represent capital recovery, operating cost, network operations, terrestrial connectivity, and safety/compliance overhead.

For definiteness, the annualized-cost values in this section are engineering normalizations rather than tariff forecasts, procurement estimates, or market-price predictions. Unless otherwise stated, the representative \$2~million/year near-Earth OGS case should be read as a transparent anchor for sensitivity analysis: recovered installed capital plus recurring operations, site lease, network-operations support, terrestrial backhaul, maintenance, safety/compliance, and automation overhead. The purpose is to expose first-order dependence on line rate, scheduled optical time, delivered-throughput efficiency, physical site count, effective site diversity, and shared-network loading.

The primary economic variable is not cost per station; it is cost per \emph{delivered} terabyte under a required availability target. A convenient first-order model is
\begin{equation}
\Qyr\,[\TB/\mathrm{yr}] \approx 164.25\times\Rline[\Gbps]\times\Hday[\mathrm{h/day}]\times\eta,
\label{eq:qyr}
\end{equation}
where $\Rline$ is the delivered user-data-equivalent line rate, $\Hday$ is average scheduled optical contact time before weather loss, and $\eta$ is the combined efficiency factor for weather, acquisition success, coding/framing overhead if not already removed from $\Rline$, maintenance, and operational losses. Thus weather is counted in $\eta$, not again in $\Hday$. The fixed-cost component is then
\begin{equation}
\costTB \approx \frac{\Cnet}{\Qyr}.
\label{eq:ctb}
\end{equation}

The numerical coefficient 164.25 in Eq.~(\ref{eq:qyr}) follows directly from $365\times3600/8/10^3$ and therefore corresponds to decimal terabytes (TB), not tebibytes (TiB).

\begin{figure}[htbp]
\centering
\includegraphics[width=0.52\linewidth]{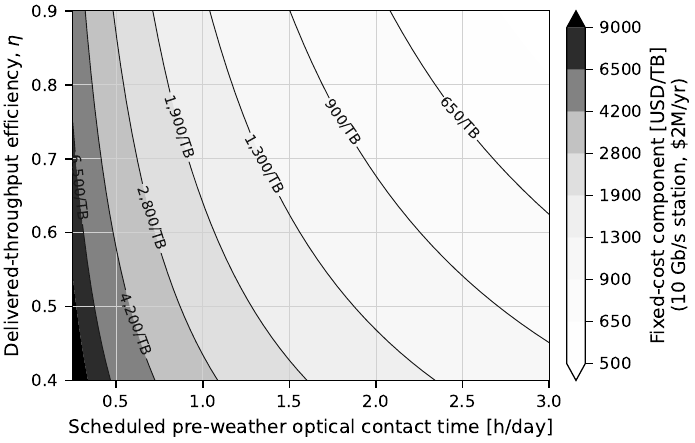}
\caption{Planning-level fixed-cost component per delivered terabyte from Eqs.~\eqref{eq:qyr}--\eqref{eq:ctb} for a representative \SI{10}{\Gbps} OGS with annualized network cost of \$2~million/year. Contours are in USD/TB and isolate sensitivity to scheduled pre-weather optical contact time and weather-inclusive delivered-throughput efficiency $\eta$. The values are scenario normalizations rather than realized service prices.}
\label{fig:costtb}
\end{figure}

For a representative \SI{10}{\Gbps} standardized near-Earth station with annualized cost \$2~million/year, scheduled pre-weather optical contact time of 0.5~h/day, and $\eta=0.7$, Eqs.~(\ref{eq:qyr}) and (\ref{eq:ctb}) give
\begin{equation}
\Qyr \approx 575\,\TB/\mathrm{yr},
\end{equation}
and
\begin{equation}
\costTB \approx 3.5\times10^{3}\,\$/\TB.
\end{equation}
This normalization sits near the conservative end of the scheduled-time range explored below. It is not a cost forecast; it is a sensitivity point showing that buffered, high-value science or Earth-observation return can be economically plausible under favorable utilization, while assured low-latency service requires additional diversity, scheduling, and operational-resilience assumptions. Figure~\ref{fig:costtb} shows the corresponding sensitivity surface for scheduled pre-weather contact time and weather-inclusive delivered-throughput efficiency.

\begin{figure}[htbp]
\centering
\includegraphics[width=0.52\linewidth]{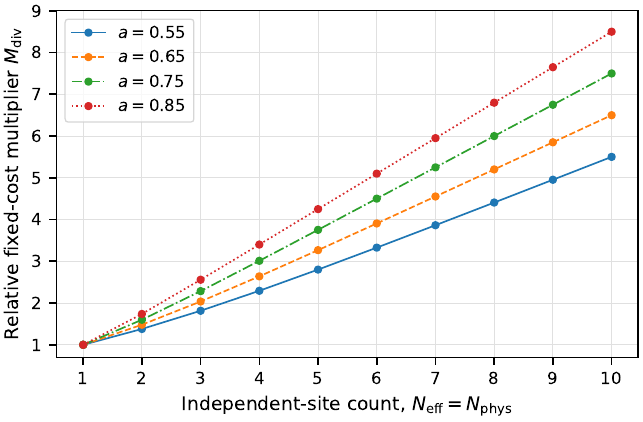}
\caption{Scenario-normalized dedicated-network fixed-cost multiplier relative to one OGS, using the independent-site limit of Eq.~\eqref{eq:multiplier} with $N_{\mathrm{eff}}=N_{\mathrm{phys}}$. Curves show $a=0.55$, 0.65, 0.75, and 0.85. They apply to the fixed-volume, availability-purchase normalization; when all physical sites are simultaneously capacity-utilized, Sec.~\ref{sec:feeder_tradeoff} gives the relevant cost model. Real networks can have $N_{\mathrm{eff}}<N_{\mathrm{phys}}$ through correlated weather, pass geometry, maintenance, scheduling, and common-mode operations.}
\label{fig:diversity}
\end{figure}

The numerical implications for representative dedicated-network cases are summarized in Table~\ref{tab:costcases}, while Fig.~\ref{fig:diversity} generalizes the same scaling over a wider range of site counts and per-site availabilities.

\subsection{Availability-adjusted cost scaling}

A dedicated mission that funds multiple physical sites does not generally receive the full independent-diversity value of those sites. Let $N_{\mathrm{phys}}$ denote the number of paid-for physical stations and $N_{\mathrm{eff}}\le N_{\mathrm{phys}}$ the effective number of statistically independent sites for the service class being analyzed. If identical sites have annualized cost $C_1$, single-site delivered throughput $Q_1$, and single-site availability $a$, then the fixed-cost multiplier for a dedicated network relative to one site is approximately
\begin{equation}
M_{\mathrm{div}}(N_{\mathrm{phys}},N_{\mathrm{eff}};a)
\approx
\frac{N_{\mathrm{phys}} a}{1-(1-a)^{N_{\mathrm{eff}}}} .
\label{eq:multiplier}
\end{equation}
The optimistic independent-site limit is recovered by setting $N_{\mathrm{eff}}=N_{\mathrm{phys}}=N$. For the illustrative case $a=0.65$, that lower-envelope limit gives $M_2\simeq1.5$, $M_3\simeq2.0$, $M_5\simeq3.3$, and $M_8\simeq5.2$. The parameters must be used consistently: in Table~\ref{tab:neff_anchor}, the DLR eight-node calibration gives $N_{\mathrm{eff}}\approx8$ if $a=0.65$, which keeps $M_{\mathrm{div}}\approx5.2$, whereas the same reported network availability gives $N_{\mathrm{eff}}=4.5$ only under the higher single-site assumption $a=0.85$, for which Eq.~\eqref{eq:multiplier} gives $M_{\mathrm{div}}\simeq6.8$. Thus $N_{\mathrm{phys}}$ drives physical cost and capacity, $N_{\mathrm{eff}}$ drives availability, and the assumed single-site availability $a$ must be held fixed within a worked example. Equation~\eqref{eq:multiplier} is normalized at fixed required delivered data volume: additional sites are purchased mainly for availability, timeliness, and continuity, not to increase all sites' simultaneously utilized capacity. If the traffic model actually fills all sites in parallel, the capacity-limited formulation in Sec.~\ref{sec:feeder_tradeoff} is the appropriate cost-per-TB model. The values below should therefore be read as scenario-normalized lower-envelope multipliers, not forecasts.

\begin{table}[htbp]
\centering
\caption{Illustrative fixed-cost burden under the representative \SI{10}{\Gbps}, \$2~million/year, 0.5~h/day pre-weather scheduled-contact, $\eta=0.7$ scenario normalization. Values use the single-site case from Eqs.~\eqref{eq:qyr}--\eqref{eq:ctb} and the optimistic independent-site limit of Eq.~\eqref{eq:multiplier}, with $N_{\mathrm{eff}}=N_{\mathrm{phys}}$ and $a=0.65$. They should be read as order-of-magnitude scenario values, not forecasts.}
\label{tab:costcases}
\setlength{\tabcolsep}{4pt}
\renewcommand{\arraystretch}{1.18}
\begin{tabular}{@{}p{0.19\textwidth}p{0.12\textwidth}p{0.17\textwidth}p{0.43\textwidth}@{}}
\toprule
Case & Relative multiplier & Delivered fixed-cost burden & Interpretation \\
\midrule
1-site case & $\sim1$ & $\sim$\$3.5k/TB & Buffered premium downlink can be plausible under favorable utilization. \\
3-site dedicated network & $\sim2$ & $\sim$\$7k/TB & Availability improves, but a single dedicated user carries most of the diversity burden. \\
5-site dedicated network & $\sim3.3$ & $\sim$\$11k/TB & More defensible for high-value or shared traffic than for lightly utilized private networks. \\
8-site dedicated network & $\sim5.2$ & $\sim$\$18k/TB & Utility-grade diversity is hard to justify without strong traffic aggregation. \\
\bottomrule
\end{tabular}
\end{table}

\subsection{Shared-network cost allocation}
\label{sec:shared_economics}

Equation~\eqref{eq:multiplier} and Table~\ref{tab:costcases} represent the unfavorable case in which one mission funds the full physical diversity network. A shared OGS network has different economics because fixed cost is allocated over aggregate traffic. Let
\begin{equation}
Q_{\mathrm{cap}}=N_{\mathrm{phys}}Q_1
\label{eq:qcap_shared}
\end{equation}
be the annual delivered-capacity normalization of $N_{\mathrm{phys}}$ physical sites under the same per-site line-rate, scheduled-time, and efficiency assumptions used in Eq.~\eqref{eq:qyr}. For customer $m$, define
\begin{equation}
u_m\equiv\frac{Q_m}{Q_{\mathrm{cap}}},\qquad
U_{\mathrm{net}}\equiv\sum_m u_m\le U_{\max}<1,
\label{eq:shared_util}
\end{equation}
where $U_{\mathrm{net}}$ is aggregate sold utilization and $U_{\max}$ is the practical utilization ceiling after scheduling guard time, weather uncertainty, maintenance, and service conflicts.

A focal customer $f$ may carry more than a proportional traffic share if it requires priority scheduling, hard-latency service, guaranteed reservation, secure integration, or mission-specific operations. A compact allocation model is
\begin{equation}
\alpha_f =
\Gamma(U_{\mathrm{net}})
\frac{w_f u_f^{\mathrm{bill}}}{\sum_m w_m u_m^{\mathrm{bill}}},
\qquad
u_m^{\mathrm{bill}}=\max(u_m,u_{\min}),
\label{eq:shared_alloc}
\end{equation}
where $w_f\ge1$ is a service-level weight, $u_{\min}$ is a minimum reserved-capacity commitment, and $\Gamma(U_{\mathrm{net}})\ge1$ is a scheduling-contention factor that increases as the network approaches $U_{\max}$.

For $C_N=N_{\mathrm{phys}}C_1+C_{\mathrm{shared}}$, the focal customer's allocated fixed-cost component per delivered terabyte is
\begin{equation}
\costTB^{\mathrm{shared}}(f)
\approx
\frac{\alpha_f C_N}{Q_f}
=
\Gamma(U_{\mathrm{net}})
\frac{w_f}{\bar w_{\mathrm{bill}}}
\frac{u_f^{\mathrm{bill}}}{u_f}
\frac{C_1}{Q_1}
\frac{1+C_{\mathrm{shared}}/(N_{\mathrm{phys}}C_1)}{U_{\mathrm{bill}}},
\label{eq:shared_ctb}
\end{equation}
with
\begin{equation}
U_{\mathrm{bill}}\equiv\sum_m u_m^{\mathrm{bill}},\qquad
\bar w_{\mathrm{bill}}\equiv
\frac{\sum_m w_m u_m^{\mathrm{bill}}}{U_{\mathrm{bill}}}.
\end{equation}
Comparing Eq.~\eqref{eq:shared_ctb} with the dedicated-user case $M_{\mathrm{div}} C_1/Q_1$ gives
\begin{equation}
\frac{\costTB^{\mathrm{shared}}(f)}
     {\costTB^{\mathrm{dedicated}}(N_{\mathrm{phys}},N_{\mathrm{eff}})}
\approx
\frac{\Gamma(U_{\mathrm{net}})}{U_{\mathrm{bill}}M_{\mathrm{div}}}
\frac{w_f}{\bar w_{\mathrm{bill}}}
\frac{u_f^{\mathrm{bill}}}{u_f}
\left(1+\frac{C_{\mathrm{shared}}}{N_{\mathrm{phys}}C_1}\right).
\label{eq:shared_ratio}
\end{equation}
Shared infrastructure is favored when aggregate utilization is high, the focal customer does not require disproportionate priority or reserved capacity, and shared overhead is small compared with the site base. For example, with $N_{\mathrm{phys}}=N_{\mathrm{eff}}=5$, $a=0.65$, $M_{\mathrm{div}}\simeq3.3$, $C_{\mathrm{shared}}/(N_{\mathrm{phys}}C_1)=0.15$, $w_f/\bar w_{\mathrm{bill}}=1.2$, $\Gamma=1.1$, and $u_f^{\mathrm{bill}}=u_f$, the shared fixed-cost burden falls below the dedicated five-site case when $U_{\mathrm{bill}}\gtrsim0.46$. This threshold is illustrative, but the scaling is general: utilization aggregation, not telescope hardware alone, is what makes weather-diverse optical service economically plausible.

\subsection{Capacity-limited versus availability-limited feeder-link economics}
\label{sec:feeder_tradeoff}

For feeder links, increasing optical line rate does not in general imply a simple inverse scaling of $\costTB$. The relevant distinction is whether the architecture is capacity-limited or availability-limited. For architecture \(x\in\{\mathrm{opt},\mathrm{RF}\}\), define the effective annual delivered throughput per site as
\begin{equation}
Q_{1,x}=164.25\,{\Rline}_x\,{\Hday}_x\,\eta_x,
\label{eq:q1x}
\end{equation}
where \({\Rline}_x\), \({\Hday}_x\), and \(\eta_x\) are the site line rate, scheduled operating hours per day, and combined efficiency factor. If \(Q_{\rm req}\) is the required annual delivered throughput and \(N_{{\rm wx},x}(A_{\rm req})\) is the minimum number of sites required to satisfy the target availability \(A_{\rm req}\), then the required site count is
\begin{equation}
N_x=\max\!\left[N_{{\rm wx},x}(A_{\rm req}),\,\left\lceil\frac{Q_{\rm req}}{Q_{1,x}}\right\rceil\right].
\label{eq:nx_feeder}
\end{equation}
The first-order fixed-cost burden then becomes
\begin{equation}
{\costTB}_x \approx \frac{N_x\,C_{{\rm site},x}+C_{{\rm shared},x}}{Q_{\rm req}},
\label{eq:ctb_feeder}
\end{equation}
where \(C_{{\rm site},x}\) is the per-site annualized cost and \(C_{{\rm shared},x}\) captures shared network costs.

Eqs.~(\ref{eq:q1x})--(\ref{eq:ctb_feeder}) imply two distinct regimes. If \(\lceil Q_{\rm req}/Q_{1,x}\rceil > N_{{\rm wx},x}\), the system is capacity-limited and \({\costTB}_x\) decreases approximately as \(1/{\Rline}_x\) at fixed \(C_{{\rm site},x}\), \({\Hday}_x\), and \(\eta_x\). If \(\lceil Q_{\rm req}/Q_{1,x}\rceil \le N_{{\rm wx},x}\), the system is availability-limited and further increases in \({\Rline}_x\) do not reduce the required site count; additional optical rate then improves margin or multiplexing headroom rather than first-order site economics. The transition occurs at the critical line rate
\begin{equation}
{\Rline^\star}_x = \frac{Q_{\rm req}}{164.25\,N_{{\rm wx},x}(A_{\rm req})\,{\Hday}_x\,\eta_x}.
\label{eq:rstar}
\end{equation}
For \({\Rline}_x>{\Rline^\star}_x\), the architecture has crossed into the availability-limited regime.

This distinction explains why higher optical feeder-link rates can materially improve economics without implying that $\costTB$ simply scales down in proportion to \(\Rline\). At the site level, a higher-rate optical gateway can have substantially lower cost per nominal \Gbps\ than an \RF\ gateway. At the network level, however, that advantage saturates once weather-diverse gateway count, rather than raw gateway capacity, becomes the dominant constraint. In that regime, the relevant economic question is not merely the cost of a faster gateway, but whether the optical architecture reduces \(N_{{\rm wx},\mathrm{opt}}(A_{\rm req})\), raises \(\eta_{\mathrm{opt}}\), or supports enough shared traffic to amortize the diversity burden.\footnote{Using purely illustrative industry values of a \SI{100}{\Gbps}-class optical gateway at EUR~3.5~million and a \SI{25}{\Gbps}-class Q/V-band \RF\ gateway at EUR~5.5~million, the site-level capital intensity per nominal \Gbps differs by a factor \((5.5/25)/(3.5/100)\approx 6.3\). This is informative only in the capacity-limited regime; once \(N_{{\rm wx},\mathrm{opt}}\) dominates, the network-level advantage depends primarily on the diversity burden rather than on raw site rate.}

\subsection{Traffic concentration and pass economics}

\begin{figure}[htbp]
\centering
\includegraphics[width=0.62\linewidth]{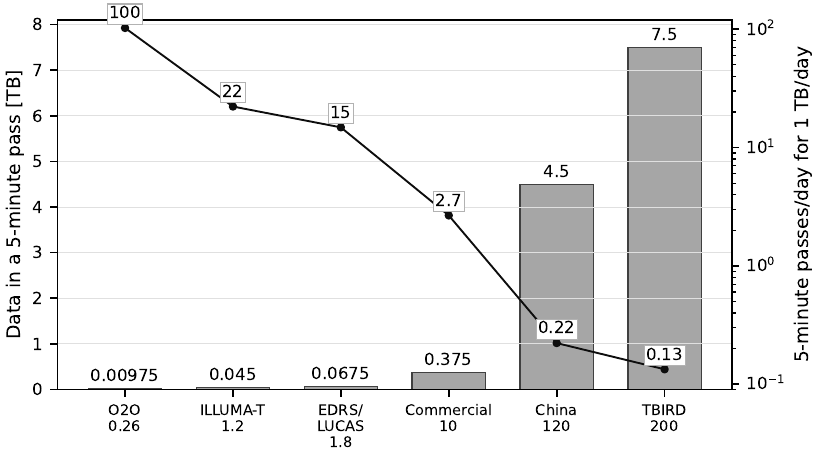}
\caption{Pass-volume scaling for selected public rate benchmarks. Bars show data returned in a \SI{5}{min} pass at the stated link rate; markers show the corresponding number of \SI{5}{min} passes per day required to deliver \SI{1}{TB/day}. This calculation isolates rate and pass concentration only; it excludes weather, acquisition loss, scheduling contention, coding overhead, backhaul, and service-level constraints.}
\label{fig:passvol}
\end{figure}

Optical service is economically most favorable when many bits can be concentrated into a short pass. For a pass duration $t_p$, the data volume is simply
\begin{equation}
V_{\mathrm{pass}} \approx \frac{R\,t_p}{8},
\label{eq:vpass}
\end{equation}
with $R$ in bit/s and $t_p$ in seconds. For a \SI{5}{min} pass, this corresponds to only \SI{9.75}{GB} at \SI{260}{\Mbps}, but \SI{67.5}{GB} at \SI{1.8}{\Gbps}, \SI{375}{GB} at \SI{10}{\Gbps}, \SI{4.5}{\TB} at \SI{120}{\Gbps}, and \SI{7.5}{\TB} at \SI{200}{\Gbps}. Figure~\ref{fig:passvol} combines this pass-volume calculation with the number of five-minute contacts per day required to deliver \SI{1}{TB/day}. The arithmetic explains why buffered \LEO\ remote-sensing missions are among the most plausible early adopters of commercial OGS service.

\subsection{Value case by mission archetype}

The planning-level value case depends on two variables: the value of incremental throughput and the required service assurance. Table~\ref{tab:usecases} summarizes the resulting market logic.

\begin{table}[htbp]
\centering
\caption{Planning-level value case by mission archetype. The case is clearest where missions can buffer data and place high value on large pass volumes.}
\label{tab:usecases}
\setlength{\tabcolsep}{4pt}
\renewcommand{\arraystretch}{1.18}
\begin{tabular}{@{}p{0.16\textwidth}p{0.19\textwidth}p{0.14\textwidth}p{0.24\textwidth}p{0.17\textwidth}@{}}
\toprule
Mission archetype & Traffic pattern & Availability requirement & Best near-term communications architecture & Business-case assessment \\
\midrule
Buffered \LEO\ Earth observation / science & High-volume burst downlink; delay tolerant & Moderate & Optical payload downlink, optionally with \RF\ backup for TT\&C & Credible current case \\
Defense mesh / ISR / proliferated transport & High aggregate traffic; resilience and low probability of intercept matter & High, but not all-weather on every path & Optical in-space mesh plus hybrid ground entry & Credible current case \\
\GEO\ relay services & Premium relay and managed service envelope & High & Optical relay with engineered ground diversity & Credible current case \\
Cislunar missions & Mixed assured operations and science return & High for command, moderate for science return & Hybrid optical + \RF & Emerging now \\
Deep-space science & Low to moderate traffic but very high science value & High mission assurance & Optical augmentation of existing \RF\ networks & Strategically relevant; limited volume \\
Telecom optical feeder links & Very high sustained throughput & Carrier-grade & Multi-site optical gateway network, likely hybridized initially & Strong future case; weaker near-term case \\
\bottomrule
\end{tabular}
\end{table}

The central trigger condition for large-scale adoption is not another rate record. It is a service inequality:
\begin{equation}
\costTB^{(\mathrm{opt})}(A_{\mathrm{req}}) < \costTB^{(\mathrm{RF})}
\quad\text{and}\quad
\Avail \ge A_{\mathrm{req}}.
\label{eq:trigger}
\end{equation}
For premium bulk downlink, Eq.~(\ref{eq:trigger}) can already be satisfied in some mission cases. For broad replacement of all-weather \RF\ gateway or TT\&C infrastructure, it is generally not yet satisfied because the network size needed to reach high $A_{\mathrm{req}}$ is still expensive unless heavily shared.

For service classes with \(T_{\max}<\Delta t_{\rm opp}\), Eq.~(\ref{eq:trigger}) should be interpreted together with Eq.~(\ref{eq:psvc}): the relevant economic question is not only the cost of annual availability, but the marginal cost of increasing decision-window success probability for the required traffic class.

\section{Public-evidence industrial map and barriers to service scaling}
\label{sec:market}

\subsection{Market layers and durable bottlenecks}

The OGS sector is best viewed as six economically distinct but linked layers: (i) demand creators, (ii) relay / in-space network providers, (iii) OGS service-network operators, (iv) OGS hardware and infrastructure suppliers, (v) atmospheric-interface and modem/baseband suppliers, and (vi) space terminal vendors. These layers have different scaling constraints and evidentiary bases. Demand creators shape the market through procurement scale and standards leverage; relay providers through installed orbital assets and service history; OGS operators through siting, automation, and customer abstraction; hardware suppliers through repeatable deployable station infrastructure; atmospheric-interface and baseband suppliers through turbulence robustness and interoperable ground electronics; and terminal vendors through qualification, reliability, and production scale. Tables~\ref{tab:scalingbarriers} and \ref{tab:valuechain} summarize the resulting value-chain structure.

A short cross-layer classification example is included in Appendix~\ref{app:cailabs} because several firms do not fit cleanly into a single value-chain layer. The Cailabs--Mbryonics comparison is used only to illustrate the classification method: Cailabs is most visible in turnkey \OGS\ hardware and atmospheric-interface photonics, whereas Mbryonics is more visible as an end-to-end optical-transport and terminal/modem supplier with \OGS\ adjacency.

\begin{table}[htbp]
\centering
\caption{Primary sources of service-layer differentiation by value-chain layer.}
\label{tab:scalingbarriers}
\setlength{\tabcolsep}{4pt}
\renewcommand{\arraystretch}{1.18}
\begin{tabular}{@{}p{0.15\textwidth}p{0.31\textwidth}p{0.21\textwidth}p{0.23\textwidth}@{}}
\toprule
Layer & Primary source of differentiation & Primary external pressure & Possible 2030 structure \\
\midrule
OGS network operators & Weather-favorable sites, remote operations, sovereign/regulatory access, cloud delivery, and customer abstraction & Large GSaaS or telecom incumbents adding optical capacity & Small number of scaled operators plus regional specialists \\
Terminal vendors & Space qualification, reliability, standards compliance, and production scale & Defense-funded or sovereign challengers with anchor procurement & Concentrated, especially in restricted sovereign markets \\
Atmospheric-interface suppliers & IP in \AO, beam combining, MPLC, detector chains, and integration know-how & In-house integration by large primes or agencies & Specialist niche with strategic leverage \\
Relay-service providers & Orbital assets, installed user base, certification, and service record & New relay constellations or large LEO mesh operators & One documented benchmark plus a few emerging challengers \\
Ground modem/ baseband suppliers & Standards support, mission-integration history, and modem interoperability & Vertical integration by terminal or OGS primes & Important enabling layer, but less concentrated than terminals \\
\bottomrule
\end{tabular}
\end{table}

\begin{table}[htbp]
\centering
\caption{Functional value chain in the OGS sector. The layers are economically linked but competitively distinct.}
\label{tab:valuechain}
\setlength{\tabcolsep}{4pt}
\renewcommand{\arraystretch}{1.18}
\begin{tabular}{@{}p{0.22\textwidth}p{0.36\textwidth}p{0.34\textwidth}@{}}
\toprule
Layer & Economic role & Representative actors in this paper \\
\midrule
Demand creators & Create procurement pull, standards pressure, and anchor utilization & SDA, NASA, ESA/EU, Chinese state ecosystem, JAXA/NEC, ISRO \\[2pt]
Relay / in-space network providers & Monetize or operationalize optical relay capacity in space & Airbus/EDRS, JAXA/NEC LUCAS, Kepler, SpaceX, Kuiper Government Solutions \\[2pt]
OGS service operators & Sell or orchestrate access to ground optical capacity & KSAT, SSC, emerging SES-linked pathway \\[2pt]
OGS hardware / infrastructure suppliers & Build deployable optical-ground stations, telescopes, and transportable systems & Cailabs, Safran, Officina Stellare, Mbryonics, BridgeComm, ASTELCO \\[2pt]
Atmospheric-interface and modem/baseband suppliers & Improve turbulence robustness and provide reusable ground electronics & Cailabs, TNO/FSO Instruments, Northrop AOA Xinetics, Safran, WORK Microwave \\[2pt]
Space terminal vendors & Supply the spacecraft-side optical terminals that determine interoperability and fleet scale & TESAT, CACI, NEC, General Atomics, Viasat, Mynaric \\
\bottomrule
\end{tabular}
\end{table}

The highest barriers to scalable service are not necessarily in telescope fabrication. Durable barriers include: access to excellent-weather sites with fiber backhaul, accumulation of closed-loop operational data, interoperability with both civil and defense optical standards, and the ability to package optical passes as an easy-to-buy service rather than as a science campaign. This helps explain why the OGS operator field is still thinner than the terminal field.

A further clarification concerns the ground-segment enablement layer. In addition to telescope and enclosure suppliers, the OGS market increasingly depends on firms that supply the integrated optical-ground electronics and control stack. WORK Microwave is one visible example: its Digital Optical Ground Station (DOG) suite is a commercial end-to-end optical ground-segment platform spanning detector, electro-optical conversion, modem, amplification, and control functions, and NASA's recent state-of-the-art review explicitly places WORK Microwave in the KSAT optical-ground ecosystem \cite{WORK_DOG_2025,NASA_SST_OGS_2025}. This layer is analytically important because a repeatable ground-electronics stack is one of the preconditions for turning OGSs from bespoke projects into scalable infrastructure.

Inclusion in the industrial map requires at least one public signal of sector relevance: (i) an operational OGS network or productized OGS service, (ii) a marketed turnkey OGS product, (iii) a publicly announced multi-year OGS contract, or (iv) a standards-relevant subsystem position that materially shapes deployable ground capability. This criterion includes firms such as Cailabs, Safran Data Systems, Officina Stellare, Mbryonics, BridgeComm, and ASTELCO, while avoiding an unstructured list of general optics or telescope suppliers. \cite{Cailabs_OGS_2025,Safran_IRIS_2025,ESA_HydRON_2025,MBRYONICS_STARGATE_2025,BridgeComm_OGT_2025,KSAT_ASTELCO_2025}

\subsection{OGS hardware manufacturers and infrastructure suppliers}
\label{sec:ogs_hardware}

The current industrial landscape is not exhausted by network operators, terminal vendors, and relay-service providers. A distinct middle layer has now emerged between telescope/platform supply and multi-site OGS operations: companies that build turnkey optical ground stations, transportable optical-ground infrastructure, or tightly coupled atmosphere-aware photonic subsystems. This layer is analytically important because it shapes the cost, repeatability, and standards compatibility of deployable OGS capacity. It also contains several firms that do not fit cleanly into the current tables. Cailabs, for example, is not only an atmospheric-interface specialist but also a turnkey OGS provider through its TILBA product line. Safran Data Systems similarly combines a complete ground-station offer with the Cortex Lasercom modem stack. Officina Stellare is best understood as an infrastructure prime for fixed and transportable OGSs rather than as a network operator. Mbryonics is an emerging vertically integrated European player spanning optical terminals, testbed facilities, and a marketed optical ground station. BridgeComm and ASTELCO belong in the hardware layer as well, though with thinner public operational evidence than established service operators. \cite{Cailabs_OGS_2025,SES_Cailabs_2025,Safran_IRIS_2025,Safran_SSC_2024,ESA_HydRON_2025,Officina_OGS_2025,MBRYONICS_STARGATE_2025,EnterpriseIreland_MBRYONICS_2025,MBRYONICS_SpaceBACN_2025,BridgeComm_OGT_2025,KSAT_ASTELCO_2025} 

Mbryonics deserves an explicit treatment because it is adjacent to, but not fully coincident with, the Cailabs position in the optical-ground value chain. Public product material presents Mbryonics not only as an \OGS\ supplier through STARGATE, but as a broader end-to-end optical transport company spanning the StarCom optical terminal, StarStream photonic modem, StarBurst optical amplifiers, and coherent optical-interconnect hardware \cite{MBRYONICS_ProductOverview_2025,MBRYONICS_STARGATE_2025}. Enterprise Ireland states that the Photon-1 manufacturing facility is sized for an initial output of 500 StarCom terminals per year, with a second facility planned for output exceeding 5{,}000 units per year \cite{EnterpriseIreland_MBRYONICS_2025,ClareEcho_MBRYONICS_2025}. Mbryonics also reports that it is the optical-terminal provider for DARPA's Space-BACN program, covering the telescope, pointing-acquisition-and-tracking chain, and optical amplifiers for a low-SWaP reconfigurable intersatellite terminal \cite{MBRYONICS_SpaceBACN_2025,DARPA_SpaceBACN_2025}. 

\begin{table}[htbp]
\centering
\caption{Cross-layer classification example for two adjacent optical-communications suppliers. The comparison illustrates value-chain placement; it is not a vendor endorsement or a claim that the firms are direct substitutes.}
\label{tab:cailabs_mbryonics}
\setlength{\tabcolsep}{4pt}
\renewcommand{\arraystretch}{1.18}
\begin{tabular}{@{}p{0.18\textwidth}p{0.36\textwidth}p{0.36\textwidth}@{}}
\toprule
Dimension & Cailabs & Mbryonics \\
\midrule
Primary value-chain locus & Turnkey \OGS\ plus atmospheric-interface differentiation & End-to-end optical transport spanning \OGS, terminal, modem, and amplifiers \\
Most visible public product emphasis & TILBA-OGS, MPLC-based turbulence mitigation, transportable optical SATCOM pathway & STARGATE \OGS, StarCom optical terminal, StarStream modem, StarBurst amplifiers \\
Visible public technical signal & \SI{80}{cm} pilot \OGS, \SI{10}{\Gbps}+ product rate, \SI{100}{\Gbps}/45-mode turbulence-mitigation publications, \(D/r_0>10\) relevance \cite{Cailabs_ICSOS_2023,Cailabs_PhotonicsWest_2023,Cailabs_ICSO_2024} & Space-BACN optical-terminal role, product stack spanning terminal and modem, coherent/high-speed optical-transport positioning \cite{MBRYONICS_ProductOverview_2025,MBRYONICS_SpaceBACN_2025} \\
Visible public industrial signal & \(>10\) \OGS\ under contract; target of 50 \OGS/yr by 2027 \cite{Cailabs_Fundraise_2025,Cailabs_EIB_2025} & 500 StarCom terminals/yr at Photon-1; \(>5{,}000\)/yr planned at Photon-2 \cite{EnterpriseIreland_MBRYONICS_2025,ClareEcho_MBRYONICS_2025} \\
Bounded analytical treatment & Cross-layer \OGS\ hardware/infrastructure and atmospheric-interface supplier & Emerging terminal/modem/end-to-end photonic-transport player with \OGS\ adjacency \\
\bottomrule
\end{tabular}
\end{table}

Table~\ref{tab:cailabs_mbryonics} is included to show why cross-layer firms should not be forced into a single ranking. Cailabs is most relevant to OGS economics where turbulence-aware receive/transmit subsystems improve the delivered-throughput factor $\eta$, reduce the effective diversity requirement $N_{\mathrm{eff}}$ for a target service class, or make each station usable for more customers. Mbryonics is more relevant where the question is terminal/modem integration, manufacturing scale, and end-to-end optical transport. Both are therefore important to the OGS ecosystem, but neither should be evaluated as if it were a current multi-site OGS service operator.

One further company should be acknowledged explicitly even if it is not included among the current OGS service operators: Mynaric. Mynaric has a long public history in laser communications hardware, including optical ground-station development, and remains strategically relevant in Europe because ESA selected it in March~2026 to develop technology for the HydRON demonstration system \cite{ESA_Mynaric_HydRON_2026,Mynaric_GS_2018}. It is therefore best treated as a high-risk but still material European optical-communications player rather than omitted from the competitive landscape.

An additional Asia-Pacific entrant should also be acknowledged explicitly: CONTEC. Public evidence supports classifying CONTEC as an emerging GSaaS/\OGS\ operator rather than a current optical-service benchmark. Safran disclosed a turnkey \SI{50}{cm} \OGS\ for CONTEC in Western Australia with data rates from \SI{100}{\Mbps} to \SI{10}{\Gbps} and support for CCSDS O3K standards; Cailabs disclosed a CCSDS-, SDA-, and QKD-capable \OGS\ order for CONTEC and noted an existing 12-station RF network; and a 2025 MoU with Odysseus Space indicates continued expansion toward optical downlink infrastructure. Public recurring optical-operations metrics remain much thinner than for KSAT or SSC, so CONTEC is better treated as an emerging entrant than as a current public-evidence benchmark \cite{Safran_Contec_2022,Cailabs_Contec_2023,Odysseus_Contec_2025}.

\begin{table}[htbp]
\centering
\caption{Representative OGS hardware and infrastructure suppliers. These firms occupy the hardware/integration layer rather than the multi-site service-operator layer.}
\label{tab:ogs_hardware}
\setlength{\tabcolsep}{4pt}
\renewcommand{\arraystretch}{1.10}
\begin{tabular}{@{}p{0.13\textwidth}p{0.17\textwidth}p{0.36\textwidth}p{0.25\textwidth}@{}}
\toprule
Company & Primary role & Public OGS-related evidence & Analytical treatment \\
\midrule
Cailabs & Turnkey \OGS\ + atmospheric photonics & TILBA-OGS is a bidirectional \OGS\ positioned at 10+~Gbps; SES announced testing of Cailabs optical ground stations for future service integration \cite{Cailabs_OGS_2025,SES_Cailabs_2025}. & Cross-layer hardware/atmospheric-interface supplier; not a multi-site service operator. \\
Safran Data Systems & Turnkey \OGS\ + modem stack & IRIS is a bidirectional \OGS\ positioned up to 10~Gbps and compatible with SDA, CCSDS, and HPE standards; Safran supplied a 50~cm IRIS \OGS\ to SSC \cite{Safran_IRIS_2025,Safran_SSC_2024}. & Systems integrator and \OGS\ manufacturer with ground-modem leverage. \\
Officina Stellare & Fixed and transportable \OGS\ infrastructure & ESA HydRON transportable-\OGS\ work and 2025 \OGS\ contract announcements for DLR/OpSTAR, EuroQCI-type work, and HydRON \cite{ESA_HydRON_2025,Officina_OGS_2025}. & Infrastructure prime; not a commercial \OGS\ network operator. \\
Mbryonics & Integrated optical transport with \OGS\ adjacency & STARGATE \OGS\ appears in product material, while current public emphasis is StarCom terminals, Space-BACN optical-terminal work, and manufacturing scale-up \cite{MBRYONICS_ProductOverview_2025,MBRYONICS_SpaceBACN_2025,EnterpriseIreland_MBRYONICS_2025}. & Adjacent terminal/modem/photonic-transport player with \OGS\ relevance. \\
BridgeComm & Fixed/mobile optical ground-terminal supplier & Public fixed and mobile optical ground-terminal product material for space and airborne lasercom links \cite{BridgeComm_OGT_2025}. & Watch-list hardware supplier; public operating metrics are thin. \\
ASTELCO Systems & Telescope / platform supplier & KSAT states that the Nemea \SI{0.5}{m} \OGS\ was designed and optimized with ASTELCO Systems \cite{KSAT_ASTELCO_2025}. & Enabling upstream telescope/platform supplier. \\
\bottomrule
\end{tabular}
\end{table}

Table~\ref{tab:ogs_hardware} separates this hardware/infrastructure layer from the service-operator layer. This distinction matters analytically: the systems question for KSAT or SSC is whether they can run a weather-diverse, automated, standards-based service network; the systems question for Cailabs, Safran, Officina Stellare, Mbryonics, BridgeComm, or ASTELCO is whether they can become preferred suppliers of repeatable OGS hardware, atmosphere-aware subsystems, or deployable station infrastructure to operators, agencies, and sovereign buyers.

Recent industry feedback suggests that the active competitor set in \OGS\ hardware may be narrower than public product pages alone imply. Consistent with the public-evidence methodology of this paper, BridgeComm is therefore treated here as historical or adjacent rather than as a current standalone \OGS\ competitor following Voyager's acquisition of BridgeComm optical communications technology in September~2025 \cite{Voyager_BridgeComm_2025}. Mynaric remains relevant to the optical communications ecosystem through terminals and HydRON-related work, but is not treated here as a current \OGS\ hardware supplier \cite{ESA_Mynaric_HydRON_2026}. Mbryonics remains relevant as an adjacent integrated photonic-transport player, but its current public emphasis appears stronger in terminals, modems, and end-to-end optical transport than in demonstrated operational \OGS\ deployment \cite{MBRYONICS_ProductOverview_2025,MBRYONICS_SpaceBACN_2025}. By contrast, General Atomics merits a visible role in the defense optical-terminal ecosystem, given its 2025 U.S. Space Force Enterprise Space Terminal Phase~2 award and its successful bi-directional air-to-space optical communications demonstration with Kepler \cite{GA_EST_2025,GA_AirSpace_2025}.

\subsection{Public-evidence map by segment}

Table~\ref{tab:marketmap} condenses the current public-evidence map by segment.

\begin{table}[htbp]
\centering
\caption{Public-evidence industrial map by segment, based on evidence available in March~2026. Labels identify visible role-specific positions within each value-chain layer; they are not proprietary performance rankings or market-share forecasts.}
\label{tab:marketmap}
\setlength{\tabcolsep}{4pt}
\renewcommand{\arraystretch}{1.10}
\begin{tabular}{@{}p{0.15\textwidth}p{0.17\textwidth}p{0.24\textwidth}p{0.38\textwidth}@{}}
\toprule
Segment & Most visible public position & Other visible / adjacent positions & Evidence basis \\
\midrule
Commercial \OGS\ networks & KSAT; SSC & SES; CONTEC & KSAT and SSC combine public optical-service products with mature ground-network operations. SES and CONTEC show visible gateway or \OGS\ build-out, but with thinner recurring optical-service metrics \cite{KSAT_Optical_2025,SSC_Optical_2025,SES_Cailabs_2025,Safran_Contec_2022,Cailabs_Contec_2023,Odysseus_Contec_2025}. \\
Optical terminals & TESAT & CACI, General Atomics, NEC, Viasat & TESAT has deep operational heritage and serial-production evidence; CACI, General Atomics, NEC, and Viasat occupy strong sovereign or defense-aligned roles in different ecosystems \cite{TESAT_Production_2024,CACI_CrossBeam_2025,GA_EST_2025,GA_AirSpace_2025,JAXA_LUCAS_2025,NEC_LUCAS_2025,Viasat_EST_2025}. \\
Atmospheric-interface technology & Cailabs & TNO / FSO Instruments, Northrop AOA Xinetics, Safran Data Systems & Cailabs has public MPLC and turnkey-\OGS\ evidence; TNO/FSO anchors a European \AO\ lineage; Northrop and Safran provide defense/integration evidence \cite{Cailabs_Lasercom_2025,TNO_FSO_2023,Northrop_Lasercom_2025,Safran_IRIS_2025}. \\
Relay-service providers & Airbus / EDRS & JAXA / NEC LUCAS, Kepler, SpaceX, Kuiper Government Solutions & Airbus/EDRS is the public operational benchmark; LUCAS is sovereign but operationally credible; Kepler is emerging; SpaceX and Kuiper are potential future integrators rather than clearly open relay utilities today \cite{Airbus_EDRS_2024,JAXA_LUCAS_2025,Kepler_Tranche1_2026,Starlink_Tech_2025,NASA_StarlinkRelay_2025,NASA_Kuiper_CSP_2022}. \\
Ground modem / baseband enablement & Work Microwave ecosystem role & Safran Cortex and integrated modem suppliers & Standards-compliant modem availability is a precondition for service abstraction; public supplier visibility remains thinner than in terminal or relay layers \cite{KSAT_Optical_2025,Safran_IRIS_2025}. \\
\bottomrule
\end{tabular}
\end{table}

A concise public-evidence reading by segment is therefore as follows. In commercial \OGS\ services, KSAT and SSC have the most visible public service-operator evidence among multi-site commercial ground-network organizations. In terminals, TESAT has extensive visible operational heritage, while CACI, General Atomics, and NEC occupy strong sovereign or defense-aligned positions in different ecosystems. In atmospheric-interface technology, Cailabs has visible public atmospheric-interface and turnkey-\OGS\ evidence, with TNO/FSO Instruments and other integration specialists representing adjacent strengths. In relay services, Airbus/EDRS is the best documented public operational benchmark.

That comparison should be interpreted as a service-layer judgment rather than as a claim about proprietary optical hardware leadership. KSAT's Nemea optical station is a half-meter aperture design developed with ASTELCO Systems and equipped with a Work Microwave baseband supporting data rates up to \SI{3}{\Gbps}; the architecture can be extended to telescopes of about \SI{1.8}{m} for lunar-class links. In ESA's EONN activity, four OGSs were integrated into the KSAT monitoring-and-control system and scheduled through the KSAT interface, with the Nemea station connected to the KSAT network operations center in Troms{\o}. This is why KSAT appears prominently in a service-operator comparison even though it is less central in discussions focused on optical terminals or relay payloads \cite{KSAT_Optical_2025,ESA_EONN_2024,NASA_SST_OGS_2025}.

\section{Demand creators through 2030 and scenario outlook}
\label{sec:demand}

\subsection{Demand creators}

Demand in this market is still dominated by governments and large institutional buyers rather than by broad commercial spot demand. Table~\ref{tab:demand} compares the organizations most likely to determine the scale and shape of the OGS market through 2030.

\begin{table}[htbp]
\centering
\caption{Public-evidence demand creators through 2030. The band is qualitative and reflects visible budget/program scale, standards leverage, and timeline credibility.}
\label{tab:demand}
\setlength{\tabcolsep}{4pt}
\renewcommand{\arraystretch}{1.18}
\begin{tabular}{@{}p{0.04\textwidth}p{0.17\textwidth}p{0.12\textwidth}p{0.39\textwidth}p{0.21\textwidth}@{}}
\toprule
\# & Organization & Band & Basis for demand pull & Affected segments \\
\midrule
1 & SDA / U.S. DoD / U.S. Space Force & Very strong & Largest visible near-term procurement engine for interoperable optical networking; Tranche~1 scale and explicit OCT standards create terminal and gateway demand simultaneously \cite{SDA_Tranche1_2025,GAO_SDA_2025,SDA_OCT_2025} & Terminals, optical mesh networking, hybrid ground entry, standards-compliant OGS services \\
2 & NASA & Very strong & Explicit transition toward commercial near-Earth relay services by 2031 and funded demonstrations with commercial providers \cite{NASA_CommercialPush_2025,NASA_CSP_2031_2025,NASA_SpaceX_CSP_2022,NASA_Kuiper_CSP_2022} & Relay services, OGS network services, cislunar / deep-space augmentation \\
3 & ESA / European Union & Strong & HydRON, ONN/EONN, Moonlight, and secure-connectivity programs create integrated demand for terminals, relay, and ground infrastructure \cite{ESA_HydRON_2025,ESA_Moonlight_2024,ESA_ONN_2026} & OGS networks, relay services, feeder-link technology, cislunar infrastructure \\
4 & Chinese state optical ecosystem & Strong & Rapid sovereign investment in high-rate direct downlink and high-altitude OGS infrastructure, though much demand may remain internally captured \cite{CAS_Pamir_2024,CAS_AIRSAT02_2026} & Domestic OGS networks, direct-to-ground links, sovereign terminal supply chains \\
5 & Japan optical ecosystem & Moderate & Operational relay heritage via LUCAS and growing alignment with future optical networking efforts \cite{JAXA_LUCAS_2025,ESA_SpaceCompass_2025} & Relay systems, sovereign terminal supply chain, future network interconnection \\
6 & Commercial telecom / constellation operators; emerging space-edge-compute concepts & Emerging & Commercial feeder-link and relay demand could become important if telecom or NGSO architectures scale. Space-edge-compute and orbital-data-center concepts are treated only as adjacent speculative demand amplifiers because public procurement and recurring-service evidence remain thin \cite{SES_Cailabs_2025,Kepler_Tranche1_2026,Starlink_Tech_2025,Turyshev2026OrbitalDataCenters} & Gateway networks, relay services, Tb/s feeder-link pathway, high-duty-cycle space-to-ground transport \\
7 & ISRO \& other emerging sovereign buyers & Adjacent / watch-list & Early but important signal that OGS capability is entering sovereign baseline planning \cite{ISRO_OGS_EOI_2025} & Sovereign OGS build-out, local integration ecosystem \\
\bottomrule
\end{tabular}
\end{table}

China is a material uncertainty in any global public-evidence map. Public sources support rapid sovereign investment in high-rate optical downlink and high-altitude OGS infrastructure, but supplier-level visibility, recurring service metrics, procurement structure, and the externally addressable portion of demand are more limited than in the U.S. and European records used elsewhere in this paper. The Chinese optical-ground ecosystem is therefore treated as a major sovereign demand and capability cluster, not as a fully comparable open commercial market segment.

The most visible public demand creators are the \SDA\ and NASA. The \SDA\ matters because it is simultaneously a budget engine, a standards setter, and a terminal-volume creator. NASA matters because it is explicitly attempting to transform optical communications from a government-owned capability into a commercial service category.

NASA's Communications Services Project now states explicitly that it is commercializing satellite relay communications for science missions in low Earth orbit and is investing in capabilities for missions launching as early as 2031. This makes NASA not merely a technology sponsor, but a direct anchor customer for the future optical-service market \cite{NASA_CSP_2026,NASA_CommercialPush_2025}.

A further emerging demand case is space-based edge compute and orbital data-center architectures \cite{Turyshev2026OrbitalDataCenters}. The cited orbital-data-center analysis is a non-peer-reviewed arXiv planning study by the present author, so it is used here only as a speculative adjacent-demand signal. These concepts are not yet demonstrated OGS demand drivers on the scale of SDA, NASA, ESA/EU, or sovereign defense programs. They are retained because their economic closure would depend partly on sustained space-to-ground transport in addition to in-orbit power, eclipse recharge, thermal rejection, utilization, and replacement cadence. In this paper they should be interpreted as possible future high-duty-cycle use cases for optical downlink, relay, and feeder-link infrastructure, not as near-term anchor customers.

\subsection{Likely scenarios through 2030}

Three scenarios appear most plausible:
\begin{enumerate}
\item \textit{Premium bulk downlink and defense networking dominate.} This is the baseline scenario. It would increase exposure of the corresponding value-chain layers, including publicly visible OGS operators, SDA-aligned terminal vendors, relay-service providers, and standards-compliant ground-segment suppliers.
\item \textit{Cislunar and relay infrastructure accelerate.} This scenario would increase exposure of relay-service providers, ESA-aligned networks, NASA commercial relay partners, and sovereign relay suppliers. It also increases the value of standardized, reusable OGS hardware and network orchestration.
\item \textit{Optical feeder links become the next scaling wave.} This is the highest-upside but least certain near-term scenario. It would increase exposure of gateway-network builders, atmospheric-interface specialists, telecom operators, and suppliers positioned for coherent / very-high-throughput feeder architectures \cite{ESA_FeederLink_2021,ESA_HydRON_2025,SES_Cailabs_2025}.
\end{enumerate}

A true market trigger would not be one more rate demonstration. It would be a procurement-plus-infrastructure event with four features: a large anchor customer, standards-based interoperability across multiple OGS operators, autonomous scheduling and cloud delivery that reduce operating cost, and enough weather-diverse sites that availability becomes a network property rather than a telescope property.

\subsection{Scenario-sensitivity matrix}

Table~\ref{tab:scenarios} converts the scenario discussion into a compact sensitivity map. The key lesson is that OGS adoption does not imply a single firm-level outcome: exposed actors depend strongly on whether the next scaling wave is defense networking, cislunar relay, or telecom feeder links.

\begin{table}[htbp]
\centering
\caption{Scenario-sensitive exposure of value-chain layers and representative public actors through 2030.}
\label{tab:scenarios}
\setlength{\tabcolsep}{4pt}
\renewcommand{\arraystretch}{1.18}
\begin{tabular}{@{}p{0.18\textwidth}p{0.11\textwidth}p{0.19\textwidth}p{0.225\textwidth}p{0.18\textwidth}@{}}
\toprule
Scenario & Probability through 2030 & Primary demand pull & Representative exposed actors & Key gating variable \\
\midrule
A: Premium bulk downlink and defense networking dominate & High & \SDA, NASA science, high-rate EO, sovereign defense users & KSAT, SSC, TESAT, CACI, General Atomics, Airbus relay services & Anchor procurement and standards interoperability \\[2pt]
B: Cislunar relay and lunar infrastructure accelerate & Medium & Artemis, Moonlight, agency/sovereign lunar missions & Airbus, ESA-aligned OGS networks, JAXA/NEC, SSC lunar ground services, NASA commercial relay partners & Mission cadence and commitment to shared lunar communications infrastructure \\[2pt]
C: Optical feeder links scale into mainstream telecom & Low-to-medium & Commercial GEO/ NGSO telecom operators and secure-connectivity programs & SES, Cailabs, TNO/FSO Instruments, KSAT, SSC, Officina Stellare, HydRON-aligned terminal suppliers & Carrier-grade availability and telco integration through atmosphere \\
\bottomrule
\end{tabular}
\end{table}

\section{Synthesis and conclusions}
\label{sec:conclusion}

The central conclusion of this paper is that OGSs have crossed the threshold from isolated demonstrations to mission-relevant communications infrastructure, but not yet to general-purpose utility-grade infrastructure. Peak line rate has been demonstrated often enough that it is no longer the central analytical question by itself. Public evidence now spans \SI{25}{\Mbps} from interplanetary range, \SI{260}{\Mbps} cislunar links, \SI{1.2}{\Gbps} ISS relay, \SI{1.8}{\Gbps} operational \GEO\ relay, \SI{120}{\Gbps}-class high-altitude \LEO\ direct downlink in China, and \SI{200}{\Gbps} from TBIRD \cite{NASA_DSOC_2024,NASA_O2O_2024,NASA_ILLUMAT_2025,Airbus_EDRS_2024,CAS_AIRSAT02_2026,NASA_TBIRD_2024}. The unresolved question is not whether optical links can work in controlled or operationally bounded settings. It is whether the industry can transform them into a shared, weather-diverse, standards-based, low-friction service layer.

Technically, the sector separates into at least four branches. The first is high-rate near-Earth downlink, where standardized OGSs and telecom-derived photonics can support a planning-level value case when utilization is high and service assurance is moderate. The second is operational relay, where Airbus/EDRS and JAXA/NEC LUCAS demonstrate that optical communication can be operated as a managed service in bounded architectures. The third is photon-efficient cislunar and deep-space communication, where performance is credible but ground economics remain mission-specific. The fourth is telecom feeder links, which have large long-term potential but remain constrained by carrier-grade availability requirements through atmosphere.

Economically, the first-order variables are utilization, duty factor, effective diversity, and cost allocation, not telescope hardware alone. A \SI{10}{\Gbps} station can support a favorable planning case for premium buffered data return under the normalization used here, but the same technology becomes much harder to justify if one mission must fund five or more physical diversity sites by itself. OGS-first architectures are therefore most plausible for buffered, high-value, non-safety-critical payload return. Hybrid optical+\RF\ architectures remain the defensible 2030 baseline for continuity-critical service classes.

The public-evidence industrial map points to a structured but still maturing field. Based on published operational and procurement evidence, KSAT and SSC have the most visible public commercial \OGS-operator positions; TESAT, CACI, General Atomics, and NEC occupy visible terminal positions in different sovereign or defense-aligned contexts; Cailabs, TNO/FSO Instruments, Safran, and related suppliers occupy distinct atmospheric-interface and ground-segment roles; and Airbus/EDRS remains the best documented public operational relay-service benchmark. These are role-specific public-evidence classifications, not forecasts of market share or proprietary rankings. On the demand side, the largest visible market-shaping organizations through 2030 remain the \SDA, NASA, and ESA/EU.

The most defensible planning baseline through 2030 is therefore hybrid. Optical will carry the highest-value throughput, relay, and spectrum-relief functions. \RF\ will continue to carry continuity, contingency, and command-assurance functions. The organizations that matter most are those that can turn this hybrid architecture into a scalable service layer with measured operational data, standards leverage, effective weather diversity, and enough shared utilization to make optical communications purchasable rather than merely demonstrable. The relevant availability metric is therefore service-class dependent: annual \(\Avail\) for buffered traffic, but decision-window success probability \(p_{\rm svc}(T_{\max})\) for latency-critical or continuity-critical traffic. For feeder links in particular, increasing optical line rate improves economics only while the architecture remains capacity-limited; once weather-diverse gateway count dominates, decision-window success and availability-adjusted diversity cost become more important than raw gateway throughput.

The conclusion of this paper would need revision under a limited set of events: a large anchor procurement that underwrites multi-site OGS infrastructure at scale; a relay or feeder-link architecture that demonstrates carrier-grade atmospheric availability at acceptable cost; or a standards-driven ecosystem shift that makes third-party optical-ground access as easy to buy as RF ground-station time. In the absence of one of these trigger conditions, the most technically and economically defensible planning baseline through 2030 remains hybrid optical+\RF, with optical first displacing RF in the highest-value throughput classes rather than in the highest-assurance continuity classes.

\section*{Acknowledgments}
The work described here was carried out at the Jet Propulsion Laboratory, California Institute of Technology, Pasadena, California, under a contract with the National Aeronautics and Space Administration.\
\textcopyright 2026. California Institute of Technology. Government sponsorship acknowledged.



\appendix

\section{Public-evidence scoring, source weighting, and sensitivity checks}
\label{app:scoring}

The company and demand-creator comparisons in Secs.~\ref{sec:market} and \ref{sec:demand} are intended to make qualitative public-evidence judgments auditable. The internal composites are ordinal indicators, not measurements of intrinsic company quality, undisclosed contract value, or investment attractiveness. In the public comparison tables they are reported as evidence bands rather than decimal scores to avoid false precision.

\subsection{Source classes and claim rules}

Table~\ref{tab:sourceclasses} defines the source classes used for public-evidence synthesis. The multiplier $q_s$ is not a mathematical truth value; it is a discipline rule that prevents a company press release from carrying the same evidentiary weight as a peer-reviewed performance paper, a standards document, an agency technical report, or a customer/operator statement with operational metrics.

\begin{table}[htbp]
\centering
\caption{Source classes used in the public-evidence synthesis. Company-primary material is useful but is deliberately assigned lower evidentiary weight unless corroborated by customers, agencies, standards activity, or operating metrics.}
\label{tab:sourceclasses}
\setlength{\tabcolsep}{4pt}
\renewcommand{\arraystretch}{1.18}
\begin{tabular}{@{}p{0.10\textwidth}p{0.25\textwidth}p{0.12\textwidth}p{0.43\textwidth}@{}}
\toprule
Class & Typical sources & $q_s$ & Claim rule \\
\midrule
E1 & Peer-reviewed papers, conference papers with technical records, CCSDS/SDA standards, agency technical reports & 1.00 & May support performance, equations, standards status, and operational interpretation. \\
E2 & Agency, customer, or operator public statements with operational metrics or procurement details & 0.90 & May support program status, operational use, service maturity, and demand-pull claims. \\
E3 & Company product literature, company technical notes, and official press releases & 0.60--0.75 & May support product existence, advertised capability, announced contracts, and strategic positioning; should not alone establish ``leadership''. \\
E4 & Trade press, industry news, and secondary summaries & 0.40--0.60 & Used only for context or chronology unless it quotes a primary document; not sufficient for major technical or ranking claims. \\
E5 & Inference from multiple public sources & case-specific & Allowed only when stated as inference and traceable to cited evidence; used mainly for scenario analysis and value-chain interpretation. \\
\bottomrule
\end{tabular}
\end{table}

For a scorecard dimension $j$, the public-evidence score is interpreted as a quality-weighted synthesis
\begin{equation}
 x_j = \frac{\sum_s q_s r_{j,s}}{\sum_s q_s},
\end{equation}
where $r_{j,s}\in[1,5]$ is the dimension-relevant rating implied by source $s$ and $q_s$ is the source-class multiplier in Table~\ref{tab:sourceclasses}. In practice, this rule is applied conservatively: an uncorroborated company source can raise the visibility of an actor but should not by itself move the actor into a top public-evidence position.

\subsection{Supplier, operator, and demand-creator scorecards}

For suppliers and operators, the baseline public-evidence composite is
\begin{equation}
\score_{\mathrm{sup}} = 0.30P + 0.25D + 0.20I + 0.15C + 0.10R,
\end{equation}
where $P$ is operational proof, $D$ is deployment or manufacturing scale, $I$ is interoperability / standards position, $C$ is customer access and procurement traction, and $R$ is relevance to the most plausible 2030 demand scenarios. For demand creators, the baseline score is
\begin{equation}
\score_{\mathrm{dem}} = 0.35B + 0.25P_r + 0.20I + 0.20T,
\end{equation}
where $B$ is visible budget or procurement scale, $P_r$ is program specificity, $I$ is ecosystem leverage, and $T$ is timeline credibility. Composite values are used only to assign the reporting bands in Table~\ref{tab:scorebands}; differences smaller than about 0.3 on the internal five-point scale are treated as effectively comparable unless the same ordering remains stable under the sensitivity cases below.

\begin{table}[htbp]
\centering
\caption{Reporting bands used for ordinal public-evidence scorecards. Composite scores are retained only as an internal audit device; Appendix~\ref{app:companies} reports bands to avoid false numerical precision.}
\label{tab:scorebands}
\setlength{\tabcolsep}{5pt}
\renewcommand{\arraystretch}{1.18}
\begin{tabular}{@{}p{0.20\textwidth}p{0.18\textwidth}p{0.50\textwidth}@{}}
\toprule
Internal composite range & Reported band & Interpretation \\
\midrule
$4.5\le S\le5.0$ & Very strong & Multiple high-quality public sources support the role-specific position. \\
$4.0\le S<4.5$ & Strong & Public evidence is substantial, but role scope or market openness is narrower. \\
$3.5\le S<4.0$ & Moderate & Credible public evidence exists, but operating metrics or scale evidence are incomplete. \\
$3.0\le S<3.5$ & Emerging & Relevant public signal exists, but execution or recurring-service evidence is still thin. \\
$S<3.0$ & Adjacent / watch-list & Adjacent capability or business model; limited current public evidence in this role. \\
\bottomrule
\end{tabular}
\end{table}

Tables~\ref{tab:sensitivity_weights} and \ref{tab:demand_sensitivity_weights} define the perturbation cases used to test whether the public-evidence classifications are robust to plausible changes in weighting.

\begin{table}[htbp]
\centering
\caption{Weight perturbations used to check whether company-position statements are robust or weight-sensitive. Each row sums to one.}
\label{tab:sensitivity_weights}
\setlength{\tabcolsep}{4pt}
\renewcommand{\arraystretch}{1.18}
\begin{tabular}{@{}p{0.25\textwidth}p{0.08\textwidth}p{0.08\textwidth}p{0.08\textwidth}p{0.08\textwidth}p{0.08\textwidth}@{}}
\toprule
Supplier/operator case & $P$ & $D$ & $I$ & $C$ & $R$ \\
\midrule
Baseline & 0.30 & 0.25 & 0.20 & 0.15 & 0.10 \\
Operational-proof heavy & 0.40 & 0.20 & 0.15 & 0.15 & 0.10 \\
Deployment-scale heavy & 0.25 & 0.35 & 0.15 & 0.15 & 0.10 \\
Standards/interoperability heavy & 0.25 & 0.20 & 0.30 & 0.15 & 0.10 \\
Customer-traction heavy & 0.25 & 0.20 & 0.15 & 0.30 & 0.10 \\
Scenario-relevance heavy & 0.25 & 0.20 & 0.15 & 0.15 & 0.25 \\
\bottomrule
\end{tabular}
\end{table}

\begin{table}[htbp]
\centering
\caption{Weight perturbations used for demand-creator sensitivity checks. Each row sums to one.}
\label{tab:demand_sensitivity_weights}
\setlength{\tabcolsep}{4pt}
\renewcommand{\arraystretch}{1.18}
\begin{tabular}{@{}p{0.30\textwidth}p{0.10\textwidth}p{0.10\textwidth}p{0.10\textwidth}p{0.10\textwidth}@{}}
\toprule
Demand-creator case & $B$ & $P_r$ & $I$ & $T$ \\
\midrule
Baseline & 0.35 & 0.25 & 0.20 & 0.20 \\
Budget/procurement-scale heavy & 0.45 & 0.20 & 0.20 & 0.15 \\
Program-specificity heavy & 0.30 & 0.35 & 0.20 & 0.15 \\
Ecosystem-leverage heavy & 0.30 & 0.20 & 0.35 & 0.15 \\
Timeline-credibility heavy & 0.30 & 0.20 & 0.15 & 0.35 \\
\bottomrule
\end{tabular}
\end{table}

\subsection{Sensitivity interpretation}

Table~\ref{tab:stability_summary} summarizes how the public-evidence statements in the main text should be interpreted after the sensitivity checks. The rule is intentionally conservative: where the ordering is weight-sensitive, the main text avoids strict ordinal winner/loser language and uses ``emerging'', ``adjacent'', ``comparable'', or ``watch-list'' terms.

\begin{table}[htbp]
\centering
\caption{Sensitivity summary for the role-based public-evidence comparisons.}
\label{tab:stability_summary}
\setlength{\tabcolsep}{4pt}
\renewcommand{\arraystretch}{1.18}
\begin{tabular}{@{}p{0.18\textwidth}p{0.33\textwidth}p{0.44\textwidth}@{}}
\toprule
Segment & Stable public-evidence finding & Weight-sensitive interpretation used in the paper \\
\midrule
Commercial OGS operators & KSAT and SSC remain the most visible commercial-operator pair under operational-proof and standards-heavy weighting. & Positions beyond the first two are treated as emerging or adjacent because service metrics and recurring optical operations are thinner. \\
Terminal vendors & TESAT has broad public operational heritage and production evidence. & CACI, NEC, General Atomics, and Viasat are separated by ecosystem and procurement context; strict global ordering is avoided outside Appendix~\ref{app:companies}. \\
Atmospheric-interface / ground segment & Cailabs has a visible public commercial MPLC/atmospheric-interface position. & TNO/FSO, Safran, and Northrop occupy different technical and sovereign/integration niches; lower-order rankings should be read as role-specific. \\
Relay services & Airbus/EDRS remains the most robust operational public benchmark across all supplier/operator weighting variants. & JAXA/NEC LUCAS is a sovereign operational benchmark rather than an open commercial service; Kepler, SpaceX, and Kuiper are scenario-sensitive challengers. \\
Demand creators & SDA, NASA, and ESA/EU remain the leading public demand-formation group under all demand-creator weighting variants. & The relative placement of Chinese state programs, Japan, commercial telecom/constellation operators, and emerging sovereign buyers depends strongly on assumptions about procurement openness and timeline. \\
\bottomrule
\end{tabular}
\end{table}

\subsection{Reproducible source inventory}

Table~\ref{tab:sourceinventory} gives the source families used to support the major claims. It is not a replacement for the bibliography; rather, it makes clear which types of sources support which kinds of claims.

\begin{table}[htbp]
\centering
\caption{Representative source inventory for the public-evidence synthesis. Full bibliographic details are given in the reference list.}
\label{tab:sourceinventory}
\setlength{\tabcolsep}{4pt}
\renewcommand{\arraystretch}{1.18}
\begin{tabular}{@{}p{0.18\textwidth}p{0.15\textwidth}p{0.60\textwidth}@{}}
\toprule
Claim family & Highest-confidence source class used & Representative sources and role in the paper \\
\midrule
Rate and operational benchmarks & E1/E2 & TBIRD, DSOC, O2O, ILLUMA-T/LCRD, EDRS, LUCAS, and AIRSAT/Pamir public records support the rate benchmark and point-technology maturity statements \cite{NASA_TBIRD_2024,Wang_TBIRD_2025,Riesing_TBIRD_2025,NASA_DSOC_2024,NASA_O2O_2024,NASA_ILLUMAT_2025,Airbus_EDRS_2024,JAXA_LUCAS_2025,NEC_LUCAS_2025,CAS_AIRSAT02_2026}. \\
Availability and site diversity & E1/E2 & NASA LCRD and DLR/Australasian availability studies are single-source public anchors for the site-specific $r_0$, two-site availability, and Australasian outage examples; they support the claim that site diversity, not line rate alone, controls service-layer maturity, but they are not treated as independent ensemble statistics \cite{LCRD_Availability_2017,DLR_AOGSN_2025}. \\
Standards and interoperability & E1/E2 & CCSDS optical standards, SDA OCT standards, and ESA ONN/EONN material support the standards/interoperability and service-layer claims \cite{CCSDS_Physical_2019,SDA_OCT_2025,ESA_ONN_2026,ESA_EONN_2024}. \\
Commercial OGS service operators & E2/E3 & KSAT and SSC public material supports the visible OGS-operator comparison; the confidence label reflects the amount of recurring operational evidence available publicly \cite{KSAT_Optical_2025,SSC_Optical_2025,SSC_NODES_2023}. \\
Relay-service evidence & E2/E3 & Airbus/EDRS and JAXA/NEC LUCAS sources support the relay-service benchmark statements; Kepler, SpaceX, and Kuiper sources are used for scenario-sensitive challenger discussion \cite{Airbus_EDRS_2024,JAXA_LUCAS_2025,NEC_LUCAS_2025,Kepler_Tranche1_2026,Starlink_Tech_2025,NASA_Kuiper_CSP_2022}. \\
Hardware and atmospheric-interface suppliers & E2/E3 & Cailabs, Safran, Officina Stellare, TNO/FSO Instruments, and related sources support the hardware/infrastructure and atmospheric-interface segmentation \cite{Cailabs_OGS_2025,Cailabs_Lasercom_2025,Safran_IRIS_2025,Safran_SSC_2024,ESA_HydRON_2025,Officina_OGS_2025,TNO_FSO_2023}. \\
Terminal-vendor positions & E2/E3 & TESAT, CACI, NEC, General Atomics, and Viasat sources support the terminal-vendor role map and confidence labels \cite{TESAT_Production_2024,CACI_CrossBeam_2025,JAXA_LUCAS_2025,NEC_LUCAS_2025,GA_EST_2025,GA_AirSpace_2025,Viasat_EST_2025}. \\
Demand creators and scenarios & E2/E5 & SDA, NASA, ESA/EU, Chinese state ecosystem, JAXA/NEC, and ISRO sources support the demand-creator and scenario analysis \cite{SDA_Tranche1_2025,GAO_SDA_2025,NASA_CSP_2026,NASA_CommercialPush_2025,ESA_HydRON_2025,ESA_Moonlight_2024,CAS_Pamir_2024,CAS_AIRSAT02_2026,ISRO_OGS_EOI_2025}. \\
Emerging / adjacent entrants & E3/E4 & CONTEC, Mbryonics, BridgeComm, Mynaric, and other adjacent actors are treated cautiously where public recurring-service evidence is thinner \cite{Safran_Contec_2022,Cailabs_Contec_2023,Odysseus_Contec_2025,MBRYONICS_ProductOverview_2025,MBRYONICS_SpaceBACN_2025,ESA_Mynaric_HydRON_2026,Voyager_BridgeComm_2025}. \\
\bottomrule
\end{tabular}
\end{table}

\section{Delivered-cost model notes}
\label{app:costmodel}

Eqs.~\eqref{eq:qyr}, \eqref{eq:ctb}, and \eqref{eq:multiplier} are intentionally first-order scenario normalizations. In Eq.~\eqref{eq:qyr}, $\Hday$ denotes scheduled pre-weather contact time and weather loss is included in $\eta$. In Eq.~\eqref{eq:multiplier}, $N_{\mathrm{phys}}$ charges the physical site base while $N_{\mathrm{eff}}$ controls the probability that at least one usable site is available. This is the fixed-delivered-volume / availability-purchase normalization; if all sites are filled with independent traffic, Eqs.~\eqref{eq:q1x}--\eqref{eq:ctb_feeder} are the more appropriate capacity formulation. The equations neglect correlation among site weather statistics, variation in pass geometry, customer burstiness, sovereign siting constraints, and second-order maintenance effects. They remain useful because they isolate the structural effect: OGS economics are governed by usable delivered bits, site diversity, and cost allocation. The more a mission can buffer data and share infrastructure, the better optical economics look. The more a customer demands always-available assured service, the more the architecture tends toward hybrid optical+\RF.

\section{Availability-adjusted network-cost multiplier}
\label{app:availability}

For identical optical sites with per-site availability $a$ and optimistic statistical independence, the network availability is
\begin{equation}
A_{\mathrm{net}}(N;a)=1-(1-a)^N.
\end{equation}
More generally, the physical number of paid sites $N_{\mathrm{phys}}$ and the effective number of statistically independent sites $N_{\mathrm{eff}}$ need not be equal. For a single dedicated mission with fixed required delivered data volume, where extra sites are purchased primarily for availability and timeliness rather than for linearly increasing consumed capacity, the fixed-cost component scales approximately as
\begin{equation}
\costTB(N_{\mathrm{phys}},N_{\mathrm{eff}};a)\propto \frac{N_{\mathrm{phys}}}{1-(1-a)^{N_{\mathrm{eff}}}} .
\end{equation}
Normalizing by the one-site case yields
\begin{equation}
F_{\mathrm{div}}(N_{\mathrm{phys}},N_{\mathrm{eff}};a)=\frac{aN_{\mathrm{phys}}}{1-(1-a)^{N_{\mathrm{eff}}}}.
\end{equation}
The independent-site expression is recovered by setting $N_{\mathrm{eff}}=N_{\mathrm{phys}}=N$. These equations are optimistic lower envelopes because real optical networks violate independence through common weather systems, maintenance clustering, shared software faults, and pass-geometry correlations. If all physical sites are simultaneously filled with revenue traffic, the capacity aggregation belongs in the denominator and the feeder-link formulation in Sec.~\ref{sec:feeder_tradeoff} should be used instead.

\section{Detailed company comparisons}
\label{app:companies}

The tables in this appendix report ordinal public-evidence bands, not measured performance scores. The internal numerical composites described in Appendix~\ref{app:scoring} are used only to make the synthesis auditable; they should not be interpreted as market share, technical superiority, investment attractiveness, or proprietary contract value.

This appendix collects the detailed company tables underlying Sec.~\ref{sec:market}. They are placed here to keep the main text analytical rather than float-heavy. The tables separate network operators, hardware/infrastructure suppliers, terminal vendors, atmospheric-interface specialists, and relay providers because companies in these categories are not directly interchangeable in the value chain.

\subsection{OGS network operators}
Publicly visible commercial optical-ground operations remain concentrated. Table~\ref{tab:ogsops} compares firms whose business model centers on selling access to OGS capacity and network operations, rather than merely supplying hardware.

\begin{table}[htbp]
\centering
\caption{Public-evidence comparison of commercial OGS operators and network integrators (March~2026). Bands are ordinal public-evidence categories; confidence decreases beyond the first two entries because the market is still forming.}
\label{tab:ogsops}
\setlength{\tabcolsep}{4pt}
\renewcommand{\arraystretch}{1.18}
\begin{tabular}{@{}p{0.01\textwidth}p{0.16\textwidth}p{0.10\textwidth}p{0.04\textwidth}p{0.59\textwidth}@{}}
\toprule
\# & Company & Band & Conf. & Public evidence / basis \\
\midrule
1 & KSAT & Very strong & High & First public commercial OGS service integrated into a mature ground-network/NOC model; \SI{0.5}{m} station designed for cost comparability with RF; strong customer abstraction and operations heritage \cite{KSAT_Optical_2025}. \\
2 & SSC & Strong & High & First two OGSs in Australia and Chile; productized service model with cloud delivery, API integration, and support for both \CCSDS\ O3K and \SDA\ OCT \cite{SSC_Optical_2025,SSC_NODES_2023}. \\
3 & SES (emerging gateway integrator) & Moderate & Med & Not yet a dedicated OGS-network operator, but a large telecom operator now testing \SI{10}{\Gbps}-class optical ground stations for future gateway integration \cite{SES_Cailabs_2025}. \\
4 & CONTEC (emerging Asia-Pacific GSaaS/\OGS\ entrant) & Emerging & Med & South Korean GSaaS operator with a public optical expansion path: Safran disclosed a turnkey \SI{50}{cm} \OGS\ in Western Australia supporting rates from \SI{100}{\Mbps} to \SI{10}{\Gbps} and CCSDS O3K standards; Cailabs disclosed a CCSDS/SDA/QKD-capable \OGS\ order and referenced an existing 12-station RF network; and a 2025 MoU with Odysseus Space targets expansion of optical downlink infrastructure \cite{Safran_Contec_2022,Cailabs_Contec_2023,Odysseus_Contec_2025}. \\
5 & Leaf Space (adjacent GSaaS watch list) & Adjacent / watch-list & Low & Strong RF ground-network business and ecosystem presence; not yet publicly visible as an optical operator, but the operating model is adjacent \cite{LeafSpace_2025}. \\
\bottomrule
\end{tabular}
\end{table}

\subsection{Terminal vendors}
The terminal market is broader and more industrialized than the operator market. Table~\ref{tab:terminals} compares vendors whose strategic position derives primarily from space-qualified optical terminals, manufacturing scale, and standards alignment.

\begin{table}[htbp]
\centering
\caption{Public-evidence comparison of optical-terminal vendors (March~2026). Bands are ordinal public-evidence categories.}
\label{tab:terminals}
\setlength{\tabcolsep}{4pt}
\renewcommand{\arraystretch}{1.18}
\begin{tabular}{@{}p{0.01\textwidth}p{0.13\textwidth}p{0.10\textwidth}p{0.07\textwidth}p{0.60\textwidth}@{}}
\toprule
\# & Vendor & Band & Conf. & Public evidence / basis \\
\midrule
1 & TESAT & Very strong & High & Deepest operational heritage in European relay systems; SCOT family in serial production and production expansion toward high unit volume \cite{TESAT_Production_2024,TESAT_Products_2024}. \\
2 & CACI & Strong & High & Strong U.S. sovereign position; CrossBeam is positioned as an American-made \SDA-compliant OCT and is tightly aligned to U.S. defense optical demand \cite{CACI_CrossBeam_2025}. \\
3 & NEC & Strong & High & Operational proof through LUCAS and strong  Japanese relay role; narrower visible export footprint than TESAT but high technical credibility \cite{JAXA_LUCAS_2025,NEC_LUCAS_2025}. \\
4 & General Atomics & Strong & Med & Fast-rising U.S. defense player with EST Phase~2 traction and air-to-space optical demonstration relevance \cite{GA_EST_2025,GA_AirSpace_2025}. \\
5 & Viasat & Moderate & Med & Visible defense-networking provider now maturing interoperable optical terminals; public on-orbit optical proof is thinner than for the ranks above \cite{Viasat_EST_2025}. \\
\bottomrule
\end{tabular}
\end{table}

\subsection{Atmospheric-interface specialists}

Table~\ref{tab:turbulence} compares firms whose main differentiation lies in turbulence mitigation, adaptive optics, photonic beam handling, or closely coupled atmosphere-aware ground subsystems.

\begin{table}[htbp]
\centering
\caption{Public-evidence comparison of atmospheric-interface / turbulence-mitigation suppliers (March~2026). Bands are ordinal public-evidence categories.}
\label{tab:turbulence}
\setlength{\tabcolsep}{4pt}
\renewcommand{\arraystretch}{1.18}
\begin{tabular}{@{}p{0.01\textwidth}p{0.15\textwidth}p{0.09\textwidth}p{0.04\textwidth}p{0.61\textwidth}@{}}
\toprule
\# & Company & Band & Conf. & Public evidence / basis \\
\midrule
1 & Cailabs & Very strong & High & Public commercial turbulence/interface proposition through MPLC-based TILBA products; explicit 10+~\Gbps positioning and validation through Kepler and SES-linked efforts \cite{Cailabs_Lasercom_2025,Cailabs_FAQ_2025}. \\
2 & TNO / FSO Instruments & Strong & High & Strong European adaptive-optics and system-engineering lineage with an industrialization path into secure-connectivity and gateway systems \cite{TNO_FSO_2023,ESA_AircraftGEO_2026}. \\
3 & Northrop Grumman AOA Xinetics & Moderate & Med & Defense-grade atmospheric-compensation and adaptive-optics capability, particularly relevant for high-consequence and airborne/terrestrial endpoints \cite{Northrop_Lasercom_2025}. \\
4 & Safran Data Systems & Moderate & Med & Strong integrated ground-segment position through IRIS OGS and Cortex Lasercom modem; strategically important as an integrator even if not a pure turbulence specialist \cite{Safran_IRIS_2025}. \\
\bottomrule
\end{tabular}
\end{table}

\subsection{Relay-service providers}

Table~\ref{tab:relay} compares organizations with substantial public evidence of operational optical relay capability or credible near-term relay competition.

\begin{table}[htbp]
\centering
\caption{Public-evidence comparison of relay-service providers and challengers (March~2026). Bands are ordinal public-evidence categories.}
\label{tab:relay}
\setlength{\tabcolsep}{4pt}
\renewcommand{\arraystretch}{1.18}
\begin{tabular}{@{}p{0.01\textwidth}p{0.16\textwidth}p{0.10\textwidth}p{0.04\textwidth}p{0.59\textwidth}@{}}
\toprule
\# & Organization & Band & Conf. & Public evidence / basis \\
\midrule
1 & Airbus / EDRS & Very strong & High & Well-documented operational commercial optical-relay position, with multi-year routine operations, reliability evidence, and public service metrics \cite{Airbus_EDRS_2024}. \\
2 & JAXA / NEC LUCAS & Strong & High & Operational sovereign relay architecture with real mission utility and strong technical credibility, even if not a broadly open third-party service \cite{JAXA_LUCAS_2025,NEC_LUCAS_2025}. \\
3 & Kepler & Moderate & Med & Strong emerging commercial-relay public evidence; first optical-relay tranche launched and aligned with HydRON-like development pathways \cite{Kepler_Tranche1_2026,ESA_HydRON_Element1_2024}. \\
4 & SpaceX / Starlink & Moderate & Med & Potential future integrator because it already operates a very large laser mesh, but less clearly positioned as a third-party relay utility \cite{Starlink_Tech_2025,NASA_StarlinkRelay_2025}. \\
5 & Kuiper Government Solutions & Emerging & Med & Strategically relevant because of NASA-funded commercial relay plans, but public execution evidence remains thinner than for the ranks above \cite{NASA_Kuiper_CSP_2022}. \\
\bottomrule
\end{tabular}
\end{table}


\section{Cross-layer classification example}
\label{app:cailabs}

This appendix records the rationale for treating Cailabs as a cross-layer optical-ground supplier rather than as a pure service-network operator. Public sources make Cailabs visible both as a turnkey \OGS\ supplier through its TILBA product family and as an atmospheric-interface supplier through MPLC-based turbulence-mitigation and beam-combining work \cite{Cailabs_OGS_2025,Cailabs_Lasercom_2025,Cailabs_FAQ_2025}. The public evidence supports a bounded classification: visible turbulence-aware optical-ground hardware and atmospheric-interface capability. It does not by itself establish full-service OGS network leadership, because recurring multi-site operations, service-level commitments, and closed-loop availability statistics are less visible publicly.

In the cost framework of Sec.~\ref{sec:economics}, the relevant technical question is whether atmosphere-aware receive and transmit subsystems increase the delivered-throughput factor $\eta$, reduce the effective diversity requirement $N_{\mathrm{eff}}$ for a target service class, or allow a shared operator to serve more missions per station. If those effects are demonstrated at network scale, they enter Eqs.~\eqref{eq:qyr}--\eqref{eq:ctb} as lower delivered fixed cost per terabyte and Eq.~\eqref{eq:shared_ctb} as improved shared-network utilization. If the improvement is limited to clear-sky link margin without changing usable contacts or scheduling efficiency, the economic effect is narrower.

\bibliographystyle{apsrev4-2}

%

\end{document}